\documentclass[journal]{IEEEtran}
\usepackage[latin9]{inputenc}
\usepackage{verbatim}
\usepackage{amsmath}
\usepackage{amssymb}
\usepackage{graphicx}
\usepackage{color}
 \definecolor{AliceBlue}{rgb}{0.94,0.97,1.00}
\definecolor{AntiqueWhite1}{rgb}{1.00,0.94,0.86}
\definecolor{AntiqueWhite2}{rgb}{0.93,0.87,0.80}
\definecolor{AntiqueWhite3}{rgb}{0.80,0.75,0.69}
\definecolor{AntiqueWhite4}{rgb}{0.55,0.51,0.47}
\definecolor{AntiqueWhite}{rgb}{0.98,0.92,0.84}
\definecolor{BlanchedAlmond}{rgb}{1.00,0.92,0.80}
\definecolor{BlueViolet}{rgb}{0.54,0.17,0.89}
\definecolor{CadetBlue1}{rgb}{0.60,0.96,1.00}
\definecolor{CadetBlue2}{rgb}{0.56,0.90,0.93}
\definecolor{CadetBlue3}{rgb}{0.48,0.77,0.80}
\definecolor{CadetBlue4}{rgb}{0.33,0.53,0.55}
\definecolor{CadetBlue}{rgb}{0.37,0.62,0.63}
\definecolor{CornflowerBlue}{rgb}{0.39,0.58,0.93}
\definecolor{DarkBlue}{rgb}{0.00,0.00,0.55}
\definecolor{DarkCyan}{rgb}{0.00,0.55,0.55}
\definecolor{DarkGoldenrod1}{rgb}{1.00,0.73,0.06}
\definecolor{DarkGoldenrod2}{rgb}{0.93,0.68,0.05}
\definecolor{DarkGoldenrod3}{rgb}{0.80,0.58,0.05}
\definecolor{DarkGoldenrod4}{rgb}{0.55,0.40,0.03}
\definecolor{DarkGoldenrod}{rgb}{0.72,0.53,0.04}
\definecolor{DarkGray}{rgb}{0.66,0.66,0.66}
\definecolor{DarkGreen}{rgb}{0.00,0.39,0.00}
\definecolor{DarkGrey}{rgb}{0.66,0.66,0.66}
\definecolor{DarkKhaki}{rgb}{0.74,0.72,0.42}
\definecolor{DarkMagenta}{rgb}{0.55,0.00,0.55}
\definecolor{DarkOliveGreen1}{rgb}{0.79,1.00,0.44}
\definecolor{DarkOliveGreen2}{rgb}{0.74,0.93,0.41}
\definecolor{DarkOliveGreen3}{rgb}{0.64,0.80,0.35}
\definecolor{DarkOliveGreen4}{rgb}{0.43,0.55,0.24}
\definecolor{DarkOliveGreen}{rgb}{0.33,0.42,0.18}
\definecolor{DarkOrange1}{rgb}{1.00,0.50,0.00}
\definecolor{DarkOrange2}{rgb}{0.93,0.46,0.00}
\definecolor{DarkOrange3}{rgb}{0.80,0.40,0.00}
\definecolor{DarkOrange4}{rgb}{0.55,0.27,0.00}
\definecolor{DarkOrange}{rgb}{1.00,0.55,0.00}
\definecolor{DarkOrchid1}{rgb}{0.75,0.24,1.00}
\definecolor{DarkOrchid2}{rgb}{0.70,0.23,0.93}
\definecolor{DarkOrchid3}{rgb}{0.60,0.20,0.80}
\definecolor{DarkOrchid4}{rgb}{0.41,0.13,0.55}
\definecolor{DarkOrchid}{rgb}{0.60,0.20,0.80}
\definecolor{DarkRed}{rgb}{0.55,0.00,0.00}
\definecolor{DarkSalmon}{rgb}{0.91,0.59,0.48}
\definecolor{DarkSeaGreen1}{rgb}{0.76,1.00,0.76}
\definecolor{DarkSeaGreen2}{rgb}{0.71,0.93,0.71}
\definecolor{DarkSeaGreen3}{rgb}{0.61,0.80,0.61}
\definecolor{DarkSeaGreen4}{rgb}{0.41,0.55,0.41}
\definecolor{DarkSeaGreen}{rgb}{0.56,0.74,0.56}
\definecolor{DarkSlateBlue}{rgb}{0.28,0.24,0.55}
\definecolor{DarkSlateGray1}{rgb}{0.59,1.00,1.00}
\definecolor{DarkSlateGray2}{rgb}{0.55,0.93,0.93}
\definecolor{DarkSlateGray3}{rgb}{0.47,0.80,0.80}
\definecolor{DarkSlateGray4}{rgb}{0.32,0.55,0.55}
\definecolor{DarkSlateGray}{rgb}{0.18,0.31,0.31}
\definecolor{DarkSlateGrey}{rgb}{0.18,0.31,0.31}
\definecolor{DarkTurquoise}{rgb}{0.00,0.81,0.82}
\definecolor{DarkViolet}{rgb}{0.58,0.00,0.83}
\definecolor{DeepPink1}{rgb}{1.00,0.08,0.58}
\definecolor{DeepPink2}{rgb}{0.93,0.07,0.54}
\definecolor{DeepPink3}{rgb}{0.80,0.06,0.46}
\definecolor{DeepPink4}{rgb}{0.55,0.04,0.31}
\definecolor{DeepPink}{rgb}{1.00,0.08,0.58}
\definecolor{DeepSkyBlue1}{rgb}{0.00,0.75,1.00}
\definecolor{DeepSkyBlue2}{rgb}{0.00,0.70,0.93}
\definecolor{DeepSkyBlue3}{rgb}{0.00,0.60,0.80}
\definecolor{DeepSkyBlue4}{rgb}{0.00,0.41,0.55}
\definecolor{DeepSkyBlue}{rgb}{0.00,0.75,1.00}
\definecolor{DimGray}{rgb}{0.41,0.41,0.41}
\definecolor{DimGrey}{rgb}{0.41,0.41,0.41}
\definecolor{DodgerBlue1}{rgb}{0.12,0.56,1.00}
\definecolor{DodgerBlue2}{rgb}{0.11,0.53,0.93}
\definecolor{DodgerBlue3}{rgb}{0.09,0.45,0.80}
\definecolor{DodgerBlue4}{rgb}{0.06,0.31,0.55}
\definecolor{DodgerBlue}{rgb}{0.12,0.56,1.00}
\definecolor{FloralWhite}{rgb}{1.00,0.98,0.94}
\definecolor{ForestGreen}{rgb}{0.13,0.55,0.13}
\definecolor{GhostWhite}{rgb}{0.97,0.97,1.00}
\definecolor{GreenYellow}{rgb}{0.68,1.00,0.18}
\definecolor{HotPink1}{rgb}{1.00,0.43,0.71}
\definecolor{HotPink2}{rgb}{0.93,0.42,0.65}
\definecolor{HotPink3}{rgb}{0.80,0.38,0.56}
\definecolor{HotPink4}{rgb}{0.55,0.23,0.38}
\definecolor{HotPink}{rgb}{1.00,0.41,0.71}
\definecolor{IndianRed1}{rgb}{1.00,0.42,0.42}
\definecolor{IndianRed2}{rgb}{0.93,0.39,0.39}
\definecolor{IndianRed3}{rgb}{0.80,0.33,0.33}
\definecolor{IndianRed4}{rgb}{0.55,0.23,0.23}
\definecolor{IndianRed}{rgb}{0.80,0.36,0.36}
\definecolor{LavenderBlush1}{rgb}{1.00,0.94,0.96}
\definecolor{LavenderBlush2}{rgb}{0.93,0.88,0.90}
\definecolor{LavenderBlush3}{rgb}{0.80,0.76,0.77}
\definecolor{LavenderBlush4}{rgb}{0.55,0.51,0.53}
\definecolor{LavenderBlush}{rgb}{1.00,0.94,0.96}
\definecolor{LawnGreen}{rgb}{0.49,0.99,0.00}
\definecolor{LemonChiffon1}{rgb}{1.00,0.98,0.80}
\definecolor{LemonChiffon2}{rgb}{0.93,0.91,0.75}
\definecolor{LemonChiffon3}{rgb}{0.80,0.79,0.65}
\definecolor{LemonChiffon4}{rgb}{0.55,0.54,0.44}
\definecolor{LemonChiffon}{rgb}{1.00,0.98,0.80}
\definecolor{LightBlue1}{rgb}{0.75,0.94,1.00}
\definecolor{LightBlue2}{rgb}{0.70,0.87,0.93}
\definecolor{LightBlue3}{rgb}{0.60,0.75,0.80}
\definecolor{LightBlue4}{rgb}{0.41,0.51,0.55}
\definecolor{LightBlue}{rgb}{0.68,0.85,0.90}
\definecolor{LightCoral}{rgb}{0.94,0.50,0.50}
\definecolor{LightCyan1}{rgb}{0.88,1.00,1.00}
\definecolor{LightCyan2}{rgb}{0.82,0.93,0.93}
\definecolor{LightCyan3}{rgb}{0.71,0.80,0.80}
\definecolor{LightCyan4}{rgb}{0.48,0.55,0.55}
\definecolor{LightCyan}{rgb}{0.88,1.00,1.00}
\definecolor{LightGoldenrod1}{rgb}{1.00,0.93,0.55}
\definecolor{LightGoldenrod2}{rgb}{0.93,0.86,0.51}
\definecolor{LightGoldenrod3}{rgb}{0.80,0.75,0.44}
\definecolor{LightGoldenrod4}{rgb}{0.55,0.51,0.30}
\definecolor{LightGoldenrodYellow}{rgb}{0.98,0.98,0.82}
\definecolor{LightGoldenrod}{rgb}{0.93,0.87,0.51}
\definecolor{LightGray}{rgb}{0.83,0.83,0.83}
\definecolor{LightGreen}{rgb}{0.56,0.93,0.56}
\definecolor{LightGrey}{rgb}{0.83,0.83,0.83}
\definecolor{LightPink1}{rgb}{1.00,0.68,0.73}
\definecolor{LightPink2}{rgb}{0.93,0.64,0.68}
\definecolor{LightPink3}{rgb}{0.80,0.55,0.58}
\definecolor{LightPink4}{rgb}{0.55,0.37,0.40}
\definecolor{LightPink}{rgb}{1.00,0.71,0.76}
\definecolor{LightSalmon1}{rgb}{1.00,0.63,0.48}
\definecolor{LightSalmon2}{rgb}{0.93,0.58,0.45}
\definecolor{LightSalmon3}{rgb}{0.80,0.51,0.38}
\definecolor{LightSalmon4}{rgb}{0.55,0.34,0.26}
\definecolor{LightSalmon}{rgb}{1.00,0.63,0.48}
\definecolor{LightSeaGreen}{rgb}{0.13,0.70,0.67}
\definecolor{LightSkyBlue1}{rgb}{0.69,0.89,1.00}
\definecolor{LightSkyBlue2}{rgb}{0.64,0.83,0.93}
\definecolor{LightSkyBlue3}{rgb}{0.55,0.71,0.80}
\definecolor{LightSkyBlue4}{rgb}{0.38,0.48,0.55}
\definecolor{LightSkyBlue}{rgb}{0.53,0.81,0.98}
\definecolor{LightSlateBlue}{rgb}{0.52,0.44,1.00}
\definecolor{LightSlateGray}{rgb}{0.47,0.53,0.60}
\definecolor{LightSlateGrey}{rgb}{0.47,0.53,0.60}
\definecolor{LightSteelBlue1}{rgb}{0.79,0.88,1.00}
\definecolor{LightSteelBlue2}{rgb}{0.74,0.82,0.93}
\definecolor{LightSteelBlue3}{rgb}{0.64,0.71,0.80}
\definecolor{LightSteelBlue4}{rgb}{0.43,0.48,0.55}
\definecolor{LightSteelBlue}{rgb}{0.69,0.77,0.87}
\definecolor{LightYellow1}{rgb}{1.00,1.00,0.88}
\definecolor{LightYellow2}{rgb}{0.93,0.93,0.82}
\definecolor{LightYellow3}{rgb}{0.80,0.80,0.71}
\definecolor{LightYellow4}{rgb}{0.55,0.55,0.48}
\definecolor{LightYellow}{rgb}{1.00,1.00,0.88}
\definecolor{LimeGreen}{rgb}{0.20,0.80,0.20}
\definecolor{MediumAquamarine}{rgb}{0.40,0.80,0.67}
\definecolor{MediumBlue}{rgb}{0.00,0.00,0.80}
\definecolor{MediumOrchid1}{rgb}{0.88,0.40,1.00}
\definecolor{MediumOrchid2}{rgb}{0.82,0.37,0.93}
\definecolor{MediumOrchid3}{rgb}{0.71,0.32,0.80}
\definecolor{MediumOrchid4}{rgb}{0.48,0.22,0.55}
\definecolor{MediumOrchid}{rgb}{0.73,0.33,0.83}
\definecolor{MediumPurple1}{rgb}{0.67,0.51,1.00}
\definecolor{MediumPurple2}{rgb}{0.62,0.47,0.93}
\definecolor{MediumPurple3}{rgb}{0.54,0.41,0.80}
\definecolor{MediumPurple4}{rgb}{0.36,0.28,0.55}
\definecolor{MediumPurple}{rgb}{0.58,0.44,0.86}
\definecolor{MediumSeaGreen}{rgb}{0.24,0.70,0.44}
\definecolor{MediumSlateBlue}{rgb}{0.48,0.41,0.93}
\definecolor{MediumSpringGreen}{rgb}{0.00,0.98,0.60}
\definecolor{MediumTurquoise}{rgb}{0.28,0.82,0.80}
\definecolor{MediumVioletRed}{rgb}{0.78,0.08,0.52}
\definecolor{MidnightBlue}{rgb}{0.10,0.10,0.44}
\definecolor{MintCream}{rgb}{0.96,1.00,0.98}
\definecolor{MistyRose1}{rgb}{1.00,0.89,0.88}
\definecolor{MistyRose2}{rgb}{0.93,0.84,0.82}
\definecolor{MistyRose3}{rgb}{0.80,0.72,0.71}
\definecolor{MistyRose4}{rgb}{0.55,0.49,0.48}
\definecolor{MistyRose}{rgb}{1.00,0.89,0.88}
\definecolor{NavajoWhite1}{rgb}{1.00,0.87,0.68}
\definecolor{NavajoWhite2}{rgb}{0.93,0.81,0.63}
\definecolor{NavajoWhite3}{rgb}{0.80,0.70,0.55}
\definecolor{NavajoWhite4}{rgb}{0.55,0.47,0.37}
\definecolor{NavajoWhite}{rgb}{1.00,0.87,0.68}
\definecolor{NavyBlue}{rgb}{0.00,0.00,0.50}
\definecolor{OldLace}{rgb}{0.99,0.96,0.90}
\definecolor{OliveDrab1}{rgb}{0.75,1.00,0.24}
\definecolor{OliveDrab2}{rgb}{0.70,0.93,0.23}
\definecolor{OliveDrab3}{rgb}{0.60,0.80,0.20}
\definecolor{OliveDrab4}{rgb}{0.41,0.55,0.13}
\definecolor{OliveDrab}{rgb}{0.42,0.56,0.14}
\definecolor{OrangeRed1}{rgb}{1.00,0.27,0.00}
\definecolor{OrangeRed2}{rgb}{0.93,0.25,0.00}
\definecolor{OrangeRed3}{rgb}{0.80,0.22,0.00}
\definecolor{OrangeRed4}{rgb}{0.55,0.15,0.00}
\definecolor{OrangeRed}{rgb}{1.00,0.27,0.00}
\definecolor{PaleGoldenrod}{rgb}{0.93,0.91,0.67}
\definecolor{PaleGreen1}{rgb}{0.60,1.00,0.60}
\definecolor{PaleGreen2}{rgb}{0.56,0.93,0.56}
\definecolor{PaleGreen3}{rgb}{0.49,0.80,0.49}
\definecolor{PaleGreen4}{rgb}{0.33,0.55,0.33}
\definecolor{PaleGreen}{rgb}{0.60,0.98,0.60}
\definecolor{PaleTurquoise1}{rgb}{0.73,1.00,1.00}
\definecolor{PaleTurquoise2}{rgb}{0.68,0.93,0.93}
\definecolor{PaleTurquoise3}{rgb}{0.59,0.80,0.80}
\definecolor{PaleTurquoise4}{rgb}{0.40,0.55,0.55}
\definecolor{PaleTurquoise}{rgb}{0.69,0.93,0.93}
\definecolor{PaleVioletRed1}{rgb}{1.00,0.51,0.67}
\definecolor{PaleVioletRed2}{rgb}{0.93,0.47,0.62}
\definecolor{PaleVioletRed3}{rgb}{0.80,0.41,0.54}
\definecolor{PaleVioletRed4}{rgb}{0.55,0.28,0.36}
\definecolor{PaleVioletRed}{rgb}{0.86,0.44,0.58}
\definecolor{PapayaWhip}{rgb}{1.00,0.94,0.84}
\definecolor{PeachPuff1}{rgb}{1.00,0.85,0.73}
\definecolor{PeachPuff2}{rgb}{0.93,0.80,0.68}
\definecolor{PeachPuff3}{rgb}{0.80,0.69,0.58}
\definecolor{PeachPuff4}{rgb}{0.55,0.47,0.40}
\definecolor{PeachPuff}{rgb}{1.00,0.85,0.73}
\definecolor{PowderBlue}{rgb}{0.69,0.88,0.90}
\definecolor{RosyBrown1}{rgb}{1.00,0.76,0.76}
\definecolor{RosyBrown2}{rgb}{0.93,0.71,0.71}
\definecolor{RosyBrown3}{rgb}{0.80,0.61,0.61}
\definecolor{RosyBrown4}{rgb}{0.55,0.41,0.41}
\definecolor{RosyBrown}{rgb}{0.74,0.56,0.56}
\definecolor{RoyalBlue1}{rgb}{0.28,0.46,1.00}
\definecolor{RoyalBlue2}{rgb}{0.26,0.43,0.93}
\definecolor{RoyalBlue3}{rgb}{0.23,0.37,0.80}
\definecolor{RoyalBlue4}{rgb}{0.15,0.25,0.55}
\definecolor{RoyalBlue}{rgb}{0.25,0.41,0.88}
\definecolor{SaddleBrown}{rgb}{0.55,0.27,0.07}
\definecolor{SandyBrown}{rgb}{0.96,0.64,0.38}
\definecolor{SeaGreen1}{rgb}{0.33,1.00,0.62}
\definecolor{SeaGreen2}{rgb}{0.31,0.93,0.58}
\definecolor{SeaGreen3}{rgb}{0.26,0.80,0.50}
\definecolor{SeaGreen4}{rgb}{0.18,0.55,0.34}
\definecolor{SeaGreen}{rgb}{0.18,0.55,0.34}
\definecolor{SkyBlue1}{rgb}{0.53,0.81,1.00}
\definecolor{SkyBlue2}{rgb}{0.49,0.75,0.93}
\definecolor{SkyBlue3}{rgb}{0.42,0.65,0.80}
\definecolor{SkyBlue4}{rgb}{0.29,0.44,0.55}
\definecolor{SkyBlue}{rgb}{0.53,0.81,0.92}
\definecolor{SlateBlue1}{rgb}{0.51,0.44,1.00}
\definecolor{SlateBlue2}{rgb}{0.48,0.40,0.93}
\definecolor{SlateBlue3}{rgb}{0.41,0.35,0.80}
\definecolor{SlateBlue4}{rgb}{0.28,0.24,0.55}
\definecolor{SlateBlue}{rgb}{0.42,0.35,0.80}
\definecolor{SlateGray1}{rgb}{0.78,0.89,1.00}
\definecolor{SlateGray2}{rgb}{0.73,0.83,0.93}
\definecolor{SlateGray3}{rgb}{0.62,0.71,0.80}
\definecolor{SlateGray4}{rgb}{0.42,0.48,0.55}
\definecolor{SlateGray}{rgb}{0.44,0.50,0.56}
\definecolor{SlateGrey}{rgb}{0.44,0.50,0.56}
\definecolor{SpringGreen1}{rgb}{0.00,1.00,0.50}
\definecolor{SpringGreen2}{rgb}{0.00,0.93,0.46}
\definecolor{SpringGreen3}{rgb}{0.00,0.80,0.40}
\definecolor{SpringGreen4}{rgb}{0.00,0.55,0.27}
\definecolor{SpringGreen}{rgb}{0.00,1.00,0.50}
\definecolor{SteelBlue1}{rgb}{0.39,0.72,1.00}
\definecolor{SteelBlue2}{rgb}{0.36,0.67,0.93}
\definecolor{SteelBlue3}{rgb}{0.31,0.58,0.80}
\definecolor{SteelBlue4}{rgb}{0.21,0.39,0.55}
\definecolor{SteelBlue}{rgb}{0.27,0.51,0.71}
\definecolor{VioletRed1}{rgb}{1.00,0.24,0.59}
\definecolor{VioletRed2}{rgb}{0.93,0.23,0.55}
\definecolor{VioletRed3}{rgb}{0.80,0.20,0.47}
\definecolor{VioletRed4}{rgb}{0.55,0.13,0.32}
\definecolor{VioletRed}{rgb}{0.82,0.13,0.56}
\definecolor{WhiteSmoke}{rgb}{0.96,0.96,0.96}
\definecolor{YellowGreen}{rgb}{0.60,0.80,0.20}
\definecolor{aliceblue}{rgb}{0.94,0.97,1.00}
\definecolor{antiquewhite}{rgb}{0.98,0.92,0.84}
\definecolor{aquamarine1}{rgb}{0.50,1.00,0.83}
\definecolor{aquamarine2}{rgb}{0.46,0.93,0.78}
\definecolor{aquamarine3}{rgb}{0.40,0.80,0.67}
\definecolor{aquamarine4}{rgb}{0.27,0.55,0.45}
\definecolor{aquamarine}{rgb}{0.50,1.00,0.83}
\definecolor{azure1}{rgb}{0.94,1.00,1.00}
\definecolor{azure2}{rgb}{0.88,0.93,0.93}
\definecolor{azure3}{rgb}{0.76,0.80,0.80}
\definecolor{azure4}{rgb}{0.51,0.55,0.55}
\definecolor{azure}{rgb}{0.94,1.00,1.00}
\definecolor{beige}{rgb}{0.96,0.96,0.86}
\definecolor{bisque1}{rgb}{1.00,0.89,0.77}
\definecolor{bisque2}{rgb}{0.93,0.84,0.72}
\definecolor{bisque3}{rgb}{0.80,0.72,0.62}
\definecolor{bisque4}{rgb}{0.55,0.49,0.42}
\definecolor{bisque}{rgb}{1.00,0.89,0.77}
\definecolor{black}{rgb}{0.00,0.00,0.00}
\definecolor{blanchedalmond}{rgb}{1.00,0.92,0.80}
\definecolor{blue1}{rgb}{0.00,0.00,1.00}
\definecolor{blue2}{rgb}{0.00,0.00,0.93}
\definecolor{blue3}{rgb}{0.00,0.00,0.80}
\definecolor{blue4}{rgb}{0.00,0.00,0.55}
\definecolor{blueviolet}{rgb}{0.54,0.17,0.89}
\definecolor{blue}{rgb}{0.00,0.00,1.00}
\definecolor{brown1}{rgb}{1.00,0.25,0.25}
\definecolor{brown2}{rgb}{0.93,0.23,0.23}
\definecolor{brown3}{rgb}{0.80,0.20,0.20}
\definecolor{brown4}{rgb}{0.55,0.14,0.14}
\definecolor{brown}{rgb}{0.65,0.16,0.16}
\definecolor{burlywood1}{rgb}{1.00,0.83,0.61}
\definecolor{burlywood2}{rgb}{0.93,0.77,0.57}
\definecolor{burlywood3}{rgb}{0.80,0.67,0.49}
\definecolor{burlywood4}{rgb}{0.55,0.45,0.33}
\definecolor{burlywood}{rgb}{0.87,0.72,0.53}
\definecolor{cadetblue}{rgb}{0.37,0.62,0.63}
\definecolor{chartreuse1}{rgb}{0.50,1.00,0.00}
\definecolor{chartreuse2}{rgb}{0.46,0.93,0.00}
\definecolor{chartreuse3}{rgb}{0.40,0.80,0.00}
\definecolor{chartreuse4}{rgb}{0.27,0.55,0.00}
\definecolor{chartreuse}{rgb}{0.50,1.00,0.00}
\definecolor{chocolate1}{rgb}{1.00,0.50,0.14}
\definecolor{chocolate2}{rgb}{0.93,0.46,0.13}
\definecolor{chocolate3}{rgb}{0.80,0.40,0.11}
\definecolor{chocolate4}{rgb}{0.55,0.27,0.07}
\definecolor{chocolate}{rgb}{0.82,0.41,0.12}
\definecolor{coral1}{rgb}{1.00,0.45,0.34}
\definecolor{coral2}{rgb}{0.93,0.42,0.31}
\definecolor{coral3}{rgb}{0.80,0.36,0.27}
\definecolor{coral4}{rgb}{0.55,0.24,0.18}
\definecolor{coral}{rgb}{1.00,0.50,0.31}
\definecolor{cornflowerblue}{rgb}{0.39,0.58,0.93}
\definecolor{cornsilk1}{rgb}{1.00,0.97,0.86}
\definecolor{cornsilk2}{rgb}{0.93,0.91,0.80}
\definecolor{cornsilk3}{rgb}{0.80,0.78,0.69}
\definecolor{cornsilk4}{rgb}{0.55,0.53,0.47}
\definecolor{cornsilk}{rgb}{1.00,0.97,0.86}
\definecolor{cyan1}{rgb}{0.00,1.00,1.00}
\definecolor{cyan2}{rgb}{0.00,0.93,0.93}
\definecolor{cyan3}{rgb}{0.00,0.80,0.80}
\definecolor{cyan4}{rgb}{0.00,0.55,0.55}
\definecolor{cyan}{rgb}{0.00,1.00,1.00}
\definecolor{darkblue}{rgb}{0.00,0.00,0.55}
\definecolor{darkcyan}{rgb}{0.00,0.55,0.55}
\definecolor{darkgoldenrod}{rgb}{0.72,0.53,0.04}
\definecolor{darkgray}{rgb}{0.66,0.66,0.66}
\definecolor{darkgreen}{rgb}{0.00,0.39,0.00}
\definecolor{darkgrey}{rgb}{0.66,0.66,0.66}
\definecolor{darkkhaki}{rgb}{0.74,0.72,0.42}
\definecolor{darkmagenta}{rgb}{0.55,0.00,0.55}
\definecolor{darkolive}{rgb}{0.33,0.42,0.18}
\definecolor{darkorange}{rgb}{1.00,0.55,0.00}
\definecolor{darkorchid}{rgb}{0.60,0.20,0.80}
\definecolor{darkred}{rgb}{0.55,0.00,0.00}
\definecolor{darksalmon}{rgb}{0.91,0.59,0.48}
\definecolor{darksea}{rgb}{0.56,0.74,0.56}
\definecolor{darkslate}{rgb}{0.18,0.31,0.31}
\definecolor{darkslate}{rgb}{0.18,0.31,0.31}
\definecolor{darkslate}{rgb}{0.28,0.24,0.55}
\definecolor{darkturquoise}{rgb}{0.00,0.81,0.82}
\definecolor{darkviolet}{rgb}{0.58,0.00,0.83}
\definecolor{deeppink}{rgb}{1.00,0.08,0.58}
\definecolor{deepsky}{rgb}{0.00,0.75,1.00}
\definecolor{dimgray}{rgb}{0.41,0.41,0.41}
\definecolor{dimgrey}{rgb}{0.41,0.41,0.41}
\definecolor{dodgerblue}{rgb}{0.12,0.56,1.00}
\definecolor{firebrick1}{rgb}{1.00,0.19,0.19}
\definecolor{firebrick2}{rgb}{0.93,0.17,0.17}
\definecolor{firebrick3}{rgb}{0.80,0.15,0.15}
\definecolor{firebrick4}{rgb}{0.55,0.10,0.10}
\definecolor{firebrick}{rgb}{0.70,0.13,0.13}
\definecolor{floralwhite}{rgb}{1.00,0.98,0.94}
\definecolor{forestgreen}{rgb}{0.13,0.55,0.13}
\definecolor{gainsboro}{rgb}{0.86,0.86,0.86}
\definecolor{ghostwhite}{rgb}{0.97,0.97,1.00}
\definecolor{gold1}{rgb}{1.00,0.84,0.00}
\definecolor{gold2}{rgb}{0.93,0.79,0.00}
\definecolor{gold3}{rgb}{0.80,0.68,0.00}
\definecolor{gold4}{rgb}{0.55,0.46,0.00}
\definecolor{goldenrod1}{rgb}{1.00,0.76,0.15}
\definecolor{goldenrod2}{rgb}{0.93,0.71,0.13}
\definecolor{goldenrod3}{rgb}{0.80,0.61,0.11}
\definecolor{goldenrod4}{rgb}{0.55,0.41,0.08}
\definecolor{goldenrod}{rgb}{0.85,0.65,0.13}
\definecolor{gold}{rgb}{1.00,0.84,0.00}
\definecolor{gray0}{rgb}{0.00,0.00,0.00}
\definecolor{gray100}{rgb}{1.00,1.00,1.00}
\definecolor{gray10}{rgb}{0.10,0.10,0.10}
\definecolor{gray11}{rgb}{0.11,0.11,0.11}
\definecolor{gray12}{rgb}{0.12,0.12,0.12}
\definecolor{gray13}{rgb}{0.13,0.13,0.13}
\definecolor{gray14}{rgb}{0.14,0.14,0.14}
\definecolor{gray15}{rgb}{0.15,0.15,0.15}
\definecolor{gray16}{rgb}{0.16,0.16,0.16}
\definecolor{gray17}{rgb}{0.17,0.17,0.17}
\definecolor{gray18}{rgb}{0.18,0.18,0.18}
\definecolor{gray19}{rgb}{0.19,0.19,0.19}
\definecolor{gray1}{rgb}{0.01,0.01,0.01}
\definecolor{gray20}{rgb}{0.20,0.20,0.20}
\definecolor{gray21}{rgb}{0.21,0.21,0.21}
\definecolor{gray22}{rgb}{0.22,0.22,0.22}
\definecolor{gray23}{rgb}{0.23,0.23,0.23}
\definecolor{gray24}{rgb}{0.24,0.24,0.24}
\definecolor{gray25}{rgb}{0.25,0.25,0.25}
\definecolor{gray26}{rgb}{0.26,0.26,0.26}
\definecolor{gray27}{rgb}{0.27,0.27,0.27}
\definecolor{gray28}{rgb}{0.28,0.28,0.28}
\definecolor{gray29}{rgb}{0.29,0.29,0.29}
\definecolor{gray2}{rgb}{0.02,0.02,0.02}
\definecolor{gray30}{rgb}{0.30,0.30,0.30}
\definecolor{gray31}{rgb}{0.31,0.31,0.31}
\definecolor{gray32}{rgb}{0.32,0.32,0.32}
\definecolor{gray33}{rgb}{0.33,0.33,0.33}
\definecolor{gray34}{rgb}{0.34,0.34,0.34}
\definecolor{gray35}{rgb}{0.35,0.35,0.35}
\definecolor{gray36}{rgb}{0.36,0.36,0.36}
\definecolor{gray37}{rgb}{0.37,0.37,0.37}
\definecolor{gray38}{rgb}{0.38,0.38,0.38}
\definecolor{gray39}{rgb}{0.39,0.39,0.39}
\definecolor{gray3}{rgb}{0.03,0.03,0.03}
\definecolor{gray40}{rgb}{0.40,0.40,0.40}
\definecolor{gray41}{rgb}{0.41,0.41,0.41}
\definecolor{gray42}{rgb}{0.42,0.42,0.42}
\definecolor{gray43}{rgb}{0.43,0.43,0.43}
\definecolor{gray44}{rgb}{0.44,0.44,0.44}
\definecolor{gray45}{rgb}{0.45,0.45,0.45}
\definecolor{gray46}{rgb}{0.46,0.46,0.46}
\definecolor{gray47}{rgb}{0.47,0.47,0.47}
\definecolor{gray48}{rgb}{0.48,0.48,0.48}
\definecolor{gray49}{rgb}{0.49,0.49,0.49}
\definecolor{gray4}{rgb}{0.04,0.04,0.04}
\definecolor{gray50}{rgb}{0.50,0.50,0.50}
\definecolor{gray51}{rgb}{0.51,0.51,0.51}
\definecolor{gray52}{rgb}{0.52,0.52,0.52}
\definecolor{gray53}{rgb}{0.53,0.53,0.53}
\definecolor{gray54}{rgb}{0.54,0.54,0.54}
\definecolor{gray55}{rgb}{0.55,0.55,0.55}
\definecolor{gray56}{rgb}{0.56,0.56,0.56}
\definecolor{gray57}{rgb}{0.57,0.57,0.57}
\definecolor{gray58}{rgb}{0.58,0.58,0.58}
\definecolor{gray59}{rgb}{0.59,0.59,0.59}
\definecolor{gray5}{rgb}{0.05,0.05,0.05}
\definecolor{gray60}{rgb}{0.60,0.60,0.60}
\definecolor{gray61}{rgb}{0.61,0.61,0.61}
\definecolor{gray62}{rgb}{0.62,0.62,0.62}
\definecolor{gray63}{rgb}{0.63,0.63,0.63}
\definecolor{gray64}{rgb}{0.64,0.64,0.64}
\definecolor{gray65}{rgb}{0.65,0.65,0.65}
\definecolor{gray66}{rgb}{0.66,0.66,0.66}
\definecolor{gray67}{rgb}{0.67,0.67,0.67}
\definecolor{gray68}{rgb}{0.68,0.68,0.68}
\definecolor{gray69}{rgb}{0.69,0.69,0.69}
\definecolor{gray6}{rgb}{0.06,0.06,0.06}
\definecolor{gray70}{rgb}{0.70,0.70,0.70}
\definecolor{gray71}{rgb}{0.71,0.71,0.71}
\definecolor{gray72}{rgb}{0.72,0.72,0.72}
\definecolor{gray73}{rgb}{0.73,0.73,0.73}
\definecolor{gray74}{rgb}{0.74,0.74,0.74}
\definecolor{gray75}{rgb}{0.75,0.75,0.75}
\definecolor{gray76}{rgb}{0.76,0.76,0.76}
\definecolor{gray77}{rgb}{0.77,0.77,0.77}
\definecolor{gray78}{rgb}{0.78,0.78,0.78}
\definecolor{gray79}{rgb}{0.79,0.79,0.79}
\definecolor{gray7}{rgb}{0.07,0.07,0.07}
\definecolor{gray80}{rgb}{0.80,0.80,0.80}
\definecolor{gray81}{rgb}{0.81,0.81,0.81}
\definecolor{gray82}{rgb}{0.82,0.82,0.82}
\definecolor{gray83}{rgb}{0.83,0.83,0.83}
\definecolor{gray84}{rgb}{0.84,0.84,0.84}
\definecolor{gray85}{rgb}{0.85,0.85,0.85}
\definecolor{gray86}{rgb}{0.86,0.86,0.86}
\definecolor{gray87}{rgb}{0.87,0.87,0.87}
\definecolor{gray88}{rgb}{0.88,0.88,0.88}
\definecolor{gray89}{rgb}{0.89,0.89,0.89}
\definecolor{gray8}{rgb}{0.08,0.08,0.08}
\definecolor{gray90}{rgb}{0.90,0.90,0.90}
\definecolor{gray91}{rgb}{0.91,0.91,0.91}
\definecolor{gray92}{rgb}{0.92,0.92,0.92}
\definecolor{gray93}{rgb}{0.93,0.93,0.93}
\definecolor{gray94}{rgb}{0.94,0.94,0.94}
\definecolor{gray95}{rgb}{0.95,0.95,0.95}
\definecolor{gray96}{rgb}{0.96,0.96,0.96}
\definecolor{gray97}{rgb}{0.97,0.97,0.97}
\definecolor{gray98}{rgb}{0.98,0.98,0.98}
\definecolor{gray99}{rgb}{0.99,0.99,0.99}
\definecolor{gray9}{rgb}{0.09,0.09,0.09}
\definecolor{gray}{rgb}{0.75,0.75,0.75}
\definecolor{green1}{rgb}{0.00,1.00,0.00}
\definecolor{green2}{rgb}{0.00,0.93,0.00}
\definecolor{green3}{rgb}{0.00,0.80,0.00}
\definecolor{green4}{rgb}{0.00,0.55,0.00}
\definecolor{greenyellow}{rgb}{0.68,1.00,0.18}
\definecolor{green}{rgb}{0.00,1.00,0.00}
\definecolor{grey0}{rgb}{0.00,0.00,0.00}
\definecolor{grey100}{rgb}{1.00,1.00,1.00}
\definecolor{grey10}{rgb}{0.10,0.10,0.10}
\definecolor{grey11}{rgb}{0.11,0.11,0.11}
\definecolor{grey12}{rgb}{0.12,0.12,0.12}
\definecolor{grey13}{rgb}{0.13,0.13,0.13}
\definecolor{grey14}{rgb}{0.14,0.14,0.14}
\definecolor{grey15}{rgb}{0.15,0.15,0.15}
\definecolor{grey16}{rgb}{0.16,0.16,0.16}
\definecolor{grey17}{rgb}{0.17,0.17,0.17}
\definecolor{grey18}{rgb}{0.18,0.18,0.18}
\definecolor{grey19}{rgb}{0.19,0.19,0.19}
\definecolor{grey1}{rgb}{0.01,0.01,0.01}
\definecolor{grey20}{rgb}{0.20,0.20,0.20}
\definecolor{grey21}{rgb}{0.21,0.21,0.21}
\definecolor{grey22}{rgb}{0.22,0.22,0.22}
\definecolor{grey23}{rgb}{0.23,0.23,0.23}
\definecolor{grey24}{rgb}{0.24,0.24,0.24}
\definecolor{grey25}{rgb}{0.25,0.25,0.25}
\definecolor{grey26}{rgb}{0.26,0.26,0.26}
\definecolor{grey27}{rgb}{0.27,0.27,0.27}
\definecolor{grey28}{rgb}{0.28,0.28,0.28}
\definecolor{grey29}{rgb}{0.29,0.29,0.29}
\definecolor{grey2}{rgb}{0.02,0.02,0.02}
\definecolor{grey30}{rgb}{0.30,0.30,0.30}
\definecolor{grey31}{rgb}{0.31,0.31,0.31}
\definecolor{grey32}{rgb}{0.32,0.32,0.32}
\definecolor{grey33}{rgb}{0.33,0.33,0.33}
\definecolor{grey34}{rgb}{0.34,0.34,0.34}
\definecolor{grey35}{rgb}{0.35,0.35,0.35}
\definecolor{grey36}{rgb}{0.36,0.36,0.36}
\definecolor{grey37}{rgb}{0.37,0.37,0.37}
\definecolor{grey38}{rgb}{0.38,0.38,0.38}
\definecolor{grey39}{rgb}{0.39,0.39,0.39}
\definecolor{grey3}{rgb}{0.03,0.03,0.03}
\definecolor{grey40}{rgb}{0.40,0.40,0.40}
\definecolor{grey41}{rgb}{0.41,0.41,0.41}
\definecolor{grey42}{rgb}{0.42,0.42,0.42}
\definecolor{grey43}{rgb}{0.43,0.43,0.43}
\definecolor{grey44}{rgb}{0.44,0.44,0.44}
\definecolor{grey45}{rgb}{0.45,0.45,0.45}
\definecolor{grey46}{rgb}{0.46,0.46,0.46}
\definecolor{grey47}{rgb}{0.47,0.47,0.47}
\definecolor{grey48}{rgb}{0.48,0.48,0.48}
\definecolor{grey49}{rgb}{0.49,0.49,0.49}
\definecolor{grey4}{rgb}{0.04,0.04,0.04}
\definecolor{grey50}{rgb}{0.50,0.50,0.50}
\definecolor{grey51}{rgb}{0.51,0.51,0.51}
\definecolor{grey52}{rgb}{0.52,0.52,0.52}
\definecolor{grey53}{rgb}{0.53,0.53,0.53}
\definecolor{grey54}{rgb}{0.54,0.54,0.54}
\definecolor{grey55}{rgb}{0.55,0.55,0.55}
\definecolor{grey56}{rgb}{0.56,0.56,0.56}
\definecolor{grey57}{rgb}{0.57,0.57,0.57}
\definecolor{grey58}{rgb}{0.58,0.58,0.58}
\definecolor{grey59}{rgb}{0.59,0.59,0.59}
\definecolor{grey5}{rgb}{0.05,0.05,0.05}
\definecolor{grey60}{rgb}{0.60,0.60,0.60}
\definecolor{grey61}{rgb}{0.61,0.61,0.61}
\definecolor{grey62}{rgb}{0.62,0.62,0.62}
\definecolor{grey63}{rgb}{0.63,0.63,0.63}
\definecolor{grey64}{rgb}{0.64,0.64,0.64}
\definecolor{grey65}{rgb}{0.65,0.65,0.65}
\definecolor{grey66}{rgb}{0.66,0.66,0.66}
\definecolor{grey67}{rgb}{0.67,0.67,0.67}
\definecolor{grey68}{rgb}{0.68,0.68,0.68}
\definecolor{grey69}{rgb}{0.69,0.69,0.69}
\definecolor{grey6}{rgb}{0.06,0.06,0.06}
\definecolor{grey70}{rgb}{0.70,0.70,0.70}
\definecolor{grey71}{rgb}{0.71,0.71,0.71}
\definecolor{grey72}{rgb}{0.72,0.72,0.72}
\definecolor{grey73}{rgb}{0.73,0.73,0.73}
\definecolor{grey74}{rgb}{0.74,0.74,0.74}
\definecolor{grey75}{rgb}{0.75,0.75,0.75}
\definecolor{grey76}{rgb}{0.76,0.76,0.76}
\definecolor{grey77}{rgb}{0.77,0.77,0.77}
\definecolor{grey78}{rgb}{0.78,0.78,0.78}
\definecolor{grey79}{rgb}{0.79,0.79,0.79}
\definecolor{grey7}{rgb}{0.07,0.07,0.07}
\definecolor{grey80}{rgb}{0.80,0.80,0.80}
\definecolor{grey81}{rgb}{0.81,0.81,0.81}
\definecolor{grey82}{rgb}{0.82,0.82,0.82}
\definecolor{grey83}{rgb}{0.83,0.83,0.83}
\definecolor{grey84}{rgb}{0.84,0.84,0.84}
\definecolor{grey85}{rgb}{0.85,0.85,0.85}
\definecolor{grey86}{rgb}{0.86,0.86,0.86}
\definecolor{grey87}{rgb}{0.87,0.87,0.87}
\definecolor{grey88}{rgb}{0.88,0.88,0.88}
\definecolor{grey89}{rgb}{0.89,0.89,0.89}
\definecolor{grey8}{rgb}{0.08,0.08,0.08}
\definecolor{grey90}{rgb}{0.90,0.90,0.90}
\definecolor{grey91}{rgb}{0.91,0.91,0.91}
\definecolor{grey92}{rgb}{0.92,0.92,0.92}
\definecolor{grey93}{rgb}{0.93,0.93,0.93}
\definecolor{grey94}{rgb}{0.94,0.94,0.94}
\definecolor{grey95}{rgb}{0.95,0.95,0.95}
\definecolor{grey96}{rgb}{0.96,0.96,0.96}
\definecolor{grey97}{rgb}{0.97,0.97,0.97}
\definecolor{grey98}{rgb}{0.98,0.98,0.98}
\definecolor{grey99}{rgb}{0.99,0.99,0.99}
\definecolor{grey9}{rgb}{0.09,0.09,0.09}
\definecolor{grey}{rgb}{0.75,0.75,0.75}
\definecolor{honeydew1}{rgb}{0.94,1.00,0.94}
\definecolor{honeydew2}{rgb}{0.88,0.93,0.88}
\definecolor{honeydew3}{rgb}{0.76,0.80,0.76}
\definecolor{honeydew4}{rgb}{0.51,0.55,0.51}
\definecolor{honeydew}{rgb}{0.94,1.00,0.94}
\definecolor{hotpink}{rgb}{1.00,0.41,0.71}
\definecolor{indianred}{rgb}{0.80,0.36,0.36}
\definecolor{ivory1}{rgb}{1.00,1.00,0.94}
\definecolor{ivory2}{rgb}{0.93,0.93,0.88}
\definecolor{ivory3}{rgb}{0.80,0.80,0.76}
\definecolor{ivory4}{rgb}{0.55,0.55,0.51}
\definecolor{ivory}{rgb}{1.00,1.00,0.94}
\definecolor{khaki1}{rgb}{1.00,0.96,0.56}
\definecolor{khaki2}{rgb}{0.93,0.90,0.52}
\definecolor{khaki3}{rgb}{0.80,0.78,0.45}
\definecolor{khaki4}{rgb}{0.55,0.53,0.31}
\definecolor{khaki}{rgb}{0.94,0.90,0.55}
\definecolor{lavenderblush}{rgb}{1.00,0.94,0.96}
\definecolor{lavender}{rgb}{0.90,0.90,0.98}
\definecolor{lawngreen}{rgb}{0.49,0.99,0.00}
\definecolor{lemonchiffon}{rgb}{1.00,0.98,0.80}
\definecolor{lightblue}{rgb}{0.68,0.85,0.90}
\definecolor{lightcoral}{rgb}{0.94,0.50,0.50}
\definecolor{lightcyan}{rgb}{0.88,1.00,1.00}
\definecolor{lightgoldenrod}{rgb}{0.93,0.87,0.51}
\definecolor{lightgoldenrod}{rgb}{0.98,0.98,0.82}
\definecolor{lightgray}{rgb}{0.83,0.83,0.83}
\definecolor{lightgreen}{rgb}{0.56,0.93,0.56}
\definecolor{lightgrey}{rgb}{0.83,0.83,0.83}
\definecolor{lightpink}{rgb}{1.00,0.71,0.76}
\definecolor{lightsalmon}{rgb}{1.00,0.63,0.48}
\definecolor{lightsea}{rgb}{0.13,0.70,0.67}
\definecolor{lightsky}{rgb}{0.53,0.81,0.98}
\definecolor{lightslate}{rgb}{0.47,0.53,0.60}
\definecolor{lightslate}{rgb}{0.47,0.53,0.60}
\definecolor{lightslate}{rgb}{0.52,0.44,1.00}
\definecolor{lightsteel}{rgb}{0.69,0.77,0.87}
\definecolor{lightyellow}{rgb}{1.00,1.00,0.88}
\definecolor{limegreen}{rgb}{0.20,0.80,0.20}
\definecolor{linen}{rgb}{0.98,0.94,0.90}
\definecolor{magenta1}{rgb}{1.00,0.00,1.00}
\definecolor{magenta2}{rgb}{0.93,0.00,0.93}
\definecolor{magenta3}{rgb}{0.80,0.00,0.80}
\definecolor{magenta4}{rgb}{0.55,0.00,0.55}
\definecolor{magenta}{rgb}{1.00,0.00,1.00}
\definecolor{maroon1}{rgb}{1.00,0.20,0.70}
\definecolor{maroon2}{rgb}{0.93,0.19,0.65}
\definecolor{maroon3}{rgb}{0.80,0.16,0.56}
\definecolor{maroon4}{rgb}{0.55,0.11,0.38}
\definecolor{maroon}{rgb}{0.69,0.19,0.38}
\definecolor{mediumaquamarine}{rgb}{0.40,0.80,0.67}
\definecolor{mediumblue}{rgb}{0.00,0.00,0.80}
\definecolor{mediumorchid}{rgb}{0.73,0.33,0.83}
\definecolor{mediumpurple}{rgb}{0.58,0.44,0.86}
\definecolor{mediumsea}{rgb}{0.24,0.70,0.44}
\definecolor{mediumslate}{rgb}{0.48,0.41,0.93}
\definecolor{mediumspring}{rgb}{0.00,0.98,0.60}
\definecolor{mediumturquoise}{rgb}{0.28,0.82,0.80}
\definecolor{mediumviolet}{rgb}{0.78,0.08,0.52}
\definecolor{midnightblue}{rgb}{0.10,0.10,0.44}
\definecolor{mintcream}{rgb}{0.96,1.00,0.98}
\definecolor{mistyrose}{rgb}{1.00,0.89,0.88}
\definecolor{moccasin}{rgb}{1.00,0.89,0.71}
\definecolor{navajowhite}{rgb}{1.00,0.87,0.68}
\definecolor{navyblue}{rgb}{0.00,0.00,0.50}
\definecolor{navy}{rgb}{0.00,0.00,0.50}
\definecolor{oldlace}{rgb}{0.99,0.96,0.90}
\definecolor{olivedrab}{rgb}{0.42,0.56,0.14}
\definecolor{orange1}{rgb}{1.00,0.65,0.00}
\definecolor{orange2}{rgb}{0.93,0.60,0.00}
\definecolor{orange3}{rgb}{0.80,0.52,0.00}
\definecolor{orange4}{rgb}{0.55,0.35,0.00}
\definecolor{orangered}{rgb}{1.00,0.27,0.00}
\definecolor{orange}{rgb}{1.00,0.65,0.00}
\definecolor{orchid1}{rgb}{1.00,0.51,0.98}
\definecolor{orchid2}{rgb}{0.93,0.48,0.91}
\definecolor{orchid3}{rgb}{0.80,0.41,0.79}
\definecolor{orchid4}{rgb}{0.55,0.28,0.54}
\definecolor{orchid}{rgb}{0.85,0.44,0.84}
\definecolor{palegoldenrod}{rgb}{0.93,0.91,0.67}
\definecolor{palegreen}{rgb}{0.60,0.98,0.60}
\definecolor{paleturquoise}{rgb}{0.69,0.93,0.93}
\definecolor{paleviolet}{rgb}{0.86,0.44,0.58}
\definecolor{papayawhip}{rgb}{1.00,0.94,0.84}
\definecolor{peachpuff}{rgb}{1.00,0.85,0.73}
\definecolor{peru}{rgb}{0.80,0.52,0.25}
\definecolor{pink1}{rgb}{1.00,0.71,0.77}
\definecolor{pink2}{rgb}{0.93,0.66,0.72}
\definecolor{pink3}{rgb}{0.80,0.57,0.62}
\definecolor{pink4}{rgb}{0.55,0.39,0.42}
\definecolor{pink}{rgb}{1.00,0.75,0.80}
\definecolor{plum1}{rgb}{1.00,0.73,1.00}
\definecolor{plum2}{rgb}{0.93,0.68,0.93}
\definecolor{plum3}{rgb}{0.80,0.59,0.80}
\definecolor{plum4}{rgb}{0.55,0.40,0.55}
\definecolor{plum}{rgb}{0.87,0.63,0.87}
\definecolor{powderblue}{rgb}{0.69,0.88,0.90}
\definecolor{purple1}{rgb}{0.61,0.19,1.00}
\definecolor{purple2}{rgb}{0.57,0.17,0.93}
\definecolor{purple3}{rgb}{0.49,0.15,0.80}
\definecolor{purple4}{rgb}{0.33,0.10,0.55}
\definecolor{purple}{rgb}{0.63,0.13,0.94}
\definecolor{red1}{rgb}{1.00,0.00,0.00}
\definecolor{red2}{rgb}{0.93,0.00,0.00}
\definecolor{red3}{rgb}{0.80,0.00,0.00}
\definecolor{red4}{rgb}{0.55,0.00,0.00}
\definecolor{red}{rgb}{1.00,0.00,0.00}
\definecolor{rosybrown}{rgb}{0.74,0.56,0.56}
\definecolor{royalblue}{rgb}{0.25,0.41,0.88}
\definecolor{saddlebrown}{rgb}{0.55,0.27,0.07}
\definecolor{salmon1}{rgb}{1.00,0.55,0.41}
\definecolor{salmon2}{rgb}{0.93,0.51,0.38}
\definecolor{salmon3}{rgb}{0.80,0.44,0.33}
\definecolor{salmon4}{rgb}{0.55,0.30,0.22}
\definecolor{salmon}{rgb}{0.98,0.50,0.45}
\definecolor{sandybrown}{rgb}{0.96,0.64,0.38}
\definecolor{seagreen}{rgb}{0.18,0.55,0.34}
\definecolor{seashell1}{rgb}{1.00,0.96,0.93}
\definecolor{seashell2}{rgb}{0.93,0.90,0.87}
\definecolor{seashell3}{rgb}{0.80,0.77,0.75}
\definecolor{seashell4}{rgb}{0.55,0.53,0.51}
\definecolor{seashell}{rgb}{1.00,0.96,0.93}
\definecolor{sienna1}{rgb}{1.00,0.51,0.28}
\definecolor{sienna2}{rgb}{0.93,0.47,0.26}
\definecolor{sienna3}{rgb}{0.80,0.41,0.22}
\definecolor{sienna4}{rgb}{0.55,0.28,0.15}
\definecolor{sienna}{rgb}{0.63,0.32,0.18}
\definecolor{skyblue}{rgb}{0.53,0.81,0.92}
\definecolor{slateblue}{rgb}{0.42,0.35,0.80}
\definecolor{slategray}{rgb}{0.44,0.50,0.56}
\definecolor{slategrey}{rgb}{0.44,0.50,0.56}
\definecolor{snow1}{rgb}{1.00,0.98,0.98}
\definecolor{snow2}{rgb}{0.93,0.91,0.91}
\definecolor{snow3}{rgb}{0.80,0.79,0.79}
\definecolor{snow4}{rgb}{0.55,0.54,0.54}
\definecolor{snow}{rgb}{1.00,0.98,0.98}
\definecolor{springgreen}{rgb}{0.00,1.00,0.50}
\definecolor{steelblue}{rgb}{0.27,0.51,0.71}
\definecolor{tan1}{rgb}{1.00,0.65,0.31}
\definecolor{tan2}{rgb}{0.93,0.60,0.29}
\definecolor{tan3}{rgb}{0.80,0.52,0.25}
\definecolor{tan4}{rgb}{0.55,0.35,0.17}
\definecolor{tan}{rgb}{0.82,0.71,0.55}
\definecolor{thistle1}{rgb}{1.00,0.88,1.00}
\definecolor{thistle2}{rgb}{0.93,0.82,0.93}
\definecolor{thistle3}{rgb}{0.80,0.71,0.80}
\definecolor{thistle4}{rgb}{0.55,0.48,0.55}
\definecolor{thistle}{rgb}{0.85,0.75,0.85}
\definecolor{tomato1}{rgb}{1.00,0.39,0.28}
\definecolor{tomato2}{rgb}{0.93,0.36,0.26}
\definecolor{tomato3}{rgb}{0.80,0.31,0.22}
\definecolor{tomato4}{rgb}{0.55,0.21,0.15}
\definecolor{tomato}{rgb}{1.00,0.39,0.28}
\definecolor{turquoise1}{rgb}{0.00,0.96,1.00}
\definecolor{turquoise2}{rgb}{0.00,0.90,0.93}
\definecolor{turquoise3}{rgb}{0.00,0.77,0.80}
\definecolor{turquoise4}{rgb}{0.00,0.53,0.55}
\definecolor{turquoise}{rgb}{0.25,0.88,0.82}
\definecolor{violetred}{rgb}{0.82,0.13,0.56}
\definecolor{violet}{rgb}{0.93,0.51,0.93}
\definecolor{wheat1}{rgb}{1.00,0.91,0.73}
\definecolor{wheat2}{rgb}{0.93,0.85,0.68}
\definecolor{wheat3}{rgb}{0.80,0.73,0.59}
\definecolor{wheat4}{rgb}{0.55,0.49,0.40}
\definecolor{wheat}{rgb}{0.96,0.87,0.70}
\definecolor{whitesmoke}{rgb}{0.96,0.96,0.96}
\definecolor{white}{rgb}{1.00,1.00,1.00}
\definecolor{yellow1}{rgb}{1.00,1.00,0.00}
\definecolor{yellow2}{rgb}{0.93,0.93,0.00}
\definecolor{yellow3}{rgb}{0.80,0.80,0.00}
\definecolor{yellow4}{rgb}{0.55,0.55,0.00}
\definecolor{yellowgreen}{rgb}{0.60,0.80,0.20}
\definecolor{yellow}{rgb}{1.00,1.00,0.00}

\makeatletter

\ifCLASSINFOpdf
\else
\fi
\usepackage{array}\usepackage{cite}
\usepackage{subfigure}

\newcommand{\w}{\omega}

\makeatother

\begin{document}

\title{Transformation Electromagnetics Devices Based on Printed-Circuit
Tensor Impedance Surfaces}

\author{Amit M.~Patel and Anthony~Grbic
\thanks{This work was partially supported by a Presidential Early Career Award
for Scientists and Engineers (FA9550-09-1-0696), a NSF Faculty Early
Career Development Award (ECCS-0747623), and by the Science, Mathematics
and Research for Transformation (S.M.A.R.T) Fellowship sponsored by
the U.S. Department of Defense (DoD) and the American Society for
Engineering Education (ASEE).%
} %
\thanks{Amit M. Patel and Anthony Grbic are with the Radiation Laboratory
in the Department of Electrical Engineering and Computer Science at
the University of Michigan-Ann Arbor. (email: amitmp@umich.edu, agrbic@umich.edu)%
}}
\maketitle
\begin{abstract}
A method for designing transformation electromagnetics devices using
tensor impedance surfaces (TISs) is presented. The method is first applied to idealized tensor impedance boundary conditions (TIBCs), and later to printed-circuit tensor impedance
surfaces (PCTISs). A PCTIS is a practical realization
of a TIBC. It consists of a tensor impedance sheet, which models a
subwavelength patterned metallic cladding, over a grounded dielectric
substrate. The method outlined in this paper allows anisotropic TIBCs
and PCTISs to be designed that support tangential wave vector distributions
and power flow directions specified by a coordinate transformation.
As an example, beam-shifting devices are designed, using TIBCs and PCTISs,
that allow a surface wave to be shifted laterally. The designs
are verified with a commercial full-wave electromagnetic solver. This work opens
new opportunities for the design and implementation of anisotropic
and inhomogeneous printed-circuit or graphene based surfaces that can guide or radiate electromagnetic
fields.\end{abstract}

\section{Introduction}

\IEEEPARstart{T}RANSFORMATION electromagnetics was first introduced
in 2006 \cite{pendry}. Since that time, it has been applied to the design of novel microwave and optical
devices such as cloaks, polarization splitters, and beam-benders \cite{pendry,Werner_TO_overview,Kundtz_TO}.
Transformation electromagnetics allows a field distribution to be
transformed from an initial configuration to a desired one via
a change of material parameters dictated by coordinate transformation. In
addition to volumetric designs, planar transformation-based devices
using transmission-line networks have been recently introduced in
\cite{Gok_Tensor_TL}, and subsequently pursued by other groups \cite{George_zedler_2D_TO,Kwon_Tensor_TL,liu_transmission_line_transformation,Selvan_sheared,Selvan_skewed}.

The need to integrate antennas and other electromagnetic devices onto the surfaces of vehicles and other platforms has driven interest in scalar, tensor, and periodic impedance surfaces in recent years. Planar leaky-wave antennas, based on scalar impedance surfaces, have been designed using sinusoidally modulated surface impedance profiles \cite{Patel_Leaky_Conference, Patel_Leaky_Journal} and tunable surface impedance profiles  \cite{Sievenpiper_forward_and_backward, Sievenpiper_beam_steering, Sievenpiper_steerable}. Great strides have been made in realizing practical printed devices such as holographic antennas, polarization controlling surfaces, and wave-guiding surfaces using the anisotropic properties of tensor impedance surfaces (TISs) \cite{Sievenpiper_Journal,Sievenpiper_pol_control,Sievenpiper_Conformal,Sievenpiper_Conformal_Advances, Sievenpiper_Conformal_Adaptive,Maci_Leaky_Circular,Gregoire_surface_waveguides}. In this paper, a method for designing transformation electromagnetics devices using TISs is presented.

\begin{figure}[ht]
\centering \subfigure[TIS supporting a $TM$ wave.]{ \includegraphics[width=3in]{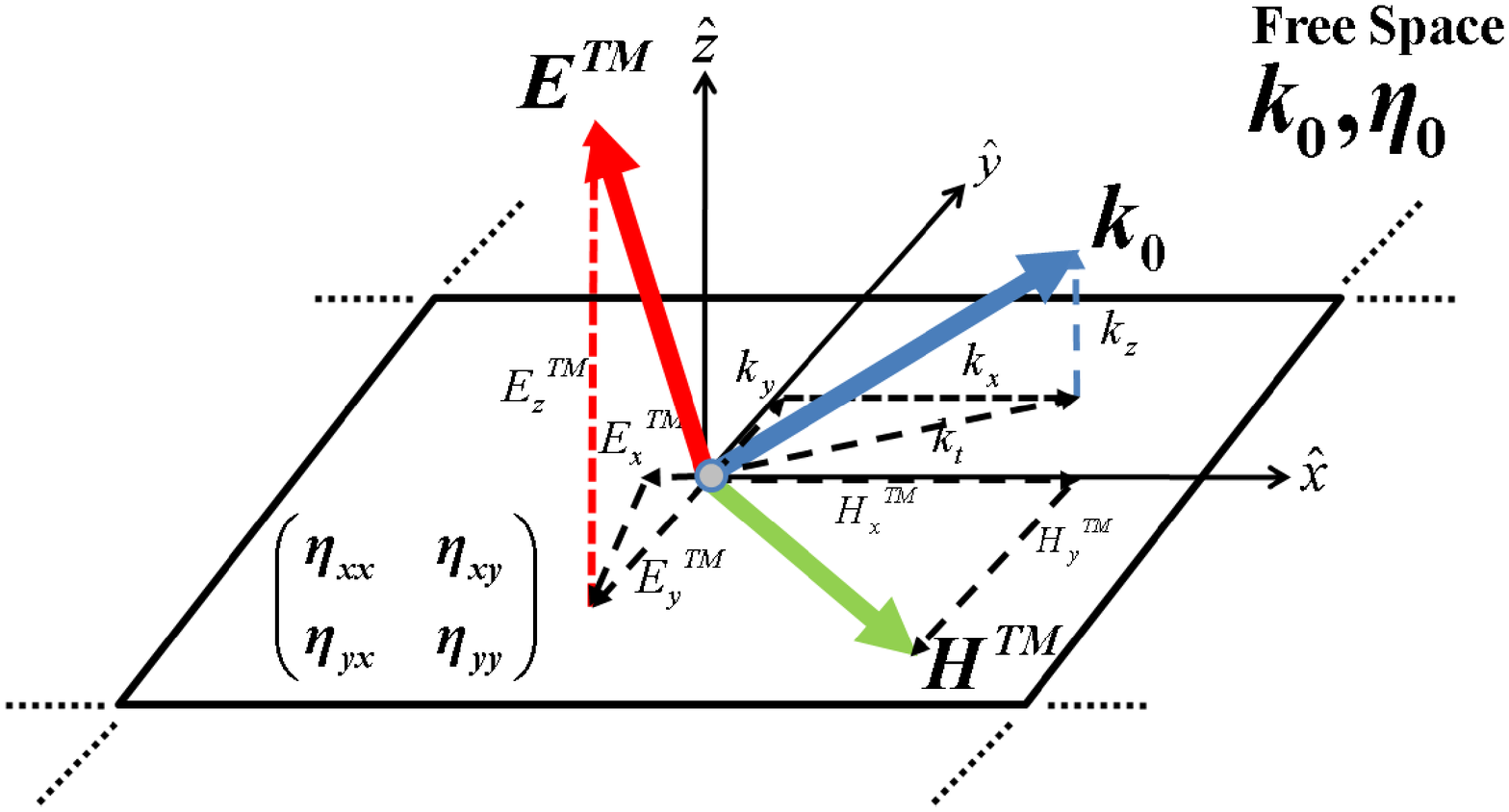}
\label{Fig:TMwave} } \subfigure[TIS supporting a $TE$ wave.]{ \includegraphics[width=3in]{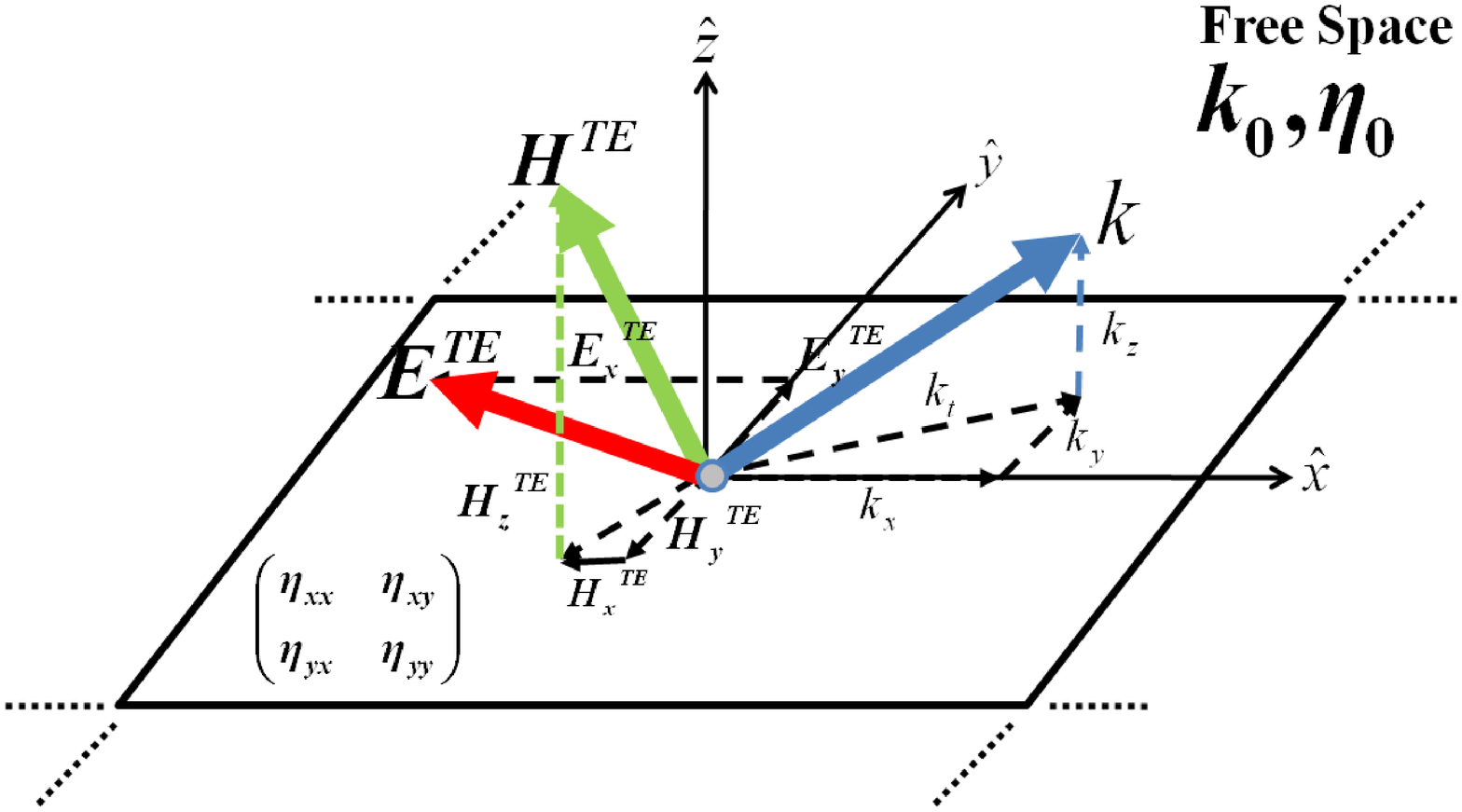}
\label{Fig:TEwave} } \caption{Waves above a tensor impedance surface (TIS). In general,
tensor impedance surfaces can support both $TM$, $TE$, and hybrid
modes. The $TM$ wave has an $E_{z}$ component and the $TE$ wave
has an $H_{z}$ component.}

\label{fig:bilow_surf}
\end{figure}

We first present a method to implement transformation
electromagnetics devices using an idealized tensor impedance boundary condition (TIBC) \cite{Bilow,Sievenpiper_Journal}.
Later in the paper, the method is adapted for printed-circuit tensor
impedance surfaces (PCTISs), which are practical realizations of TIBCs \cite{Patel_Grbic_Analytical_journal,Patel_IMS2012,Patel_APS2012,Patel_Grbic_Effective_journal}. The TIBC is given by: $\bar{E_{t}}=\bar{\bar{\eta}}_{surf}\hat{n}\times\bar{H_{t}}$,
where $\bar{E_{t}}$ and $\bar{H_{t}}$ are components of the total
electric and magnetic field tangential to the surface (at $z=0)$ and $\hat{n}$ is the surface normal\cite{Bilow}.
This boundary condition can be represented in matrix form as
\begin{equation}
\begin{pmatrix}E_{x}\\
E_{y}
\end{pmatrix}=\bar{\bar{\eta}}_{surf}\begin{pmatrix}-H_{y}\\
H_{x}
\end{pmatrix}=\begin{pmatrix}\eta_{xx} & \eta_{xy}\\
\eta_{yx} & \eta_{yy}
\end{pmatrix}\begin{pmatrix}-H_{y}\\
H_{x}
\end{pmatrix},\label{eq: TIS boundary condition}
\end{equation}
or in terms of the surface admittances as
\begin{equation}
\begin{pmatrix}-H_{y}\\
H_{x}
\end{pmatrix}=\bar{\bar{Y}}_{surf}\begin{pmatrix}E_{x}\\
E_{y}
\end{pmatrix}=\begin{pmatrix}Y_{xx} & Y_{xy}\\
Y_{yx} & Y_{yy}
\end{pmatrix}\begin{pmatrix}E_{x}\\
E_{y}
\end{pmatrix},\label{eq: Y boundary condition}
\end{equation}
where
\begin{equation}
\bar{\bar{Y}}_{surf}=\bar{\bar{\eta}}_{surf}^{-1}\label{eq: Y inverse eta}
\end{equation}
A TIS can support $TM$ (Fig. \ref{Fig:TMwave}),
$TE$ (Fig. \ref{Fig:TEwave}), or hybrid modes. Recently, a surface
impedance cloak was designed \cite{Quarfoth_APS_2012_cloak} using
the $TM$ index profile characteristic of a beam-shifter \cite{Kong_beamshifter}. In the present
work, the surface impedance profile needed to implement transformation
electromagnetics devices is found from the transformed wave vector
and Poynting vector distributions along a surface \cite{Patel_IMS2013_transformation}. Specifically,
surface impedance profiles are found that support modes ($TM$, $TE$,
or hybrid) with these transformed phase and power characteristics. The method ensures
that only the surface impedance entries need to be transformed, and the
free space above the TIS need not be transformed. The method is later
adapted to design practical PCTISs that also support modes with transformed wave
vector and Poynting vector distributions. A PCTIS is a practical realization
of a TIBC, consisting of a patterned metallic cladding over a grounded dielectric substrate. The patterned metallic cladding is modeled as a tensor sheet impedance \cite{Patel_Grbic_Analytical_journal,Patel_IMS2012,Patel_APS2012,Patel_Grbic_Effective_journal}. When designing PCTISs, the tensor sheet impedance entries
are the unknowns: the quantities of interest.

In the next section of this paper, transformation electromagnetics
in two dimensions (2D) is reviewed. Section III outlines an approach
for designing 2D transformation electromagnetics devices using TIBCs.
In Section IV, a beam-shifting device is designed and simulated with
a commercial full-wave solver to verify the design method outlined
in Section III. In Section V, transformation electromagnetics is applied
to PCTISs, and a beam-shifter is designed using a PCTIS in Section VI. The proposed
design methodology is a step towards the realization of practical,
transformation electromagnetics devices using PCTISs \cite{Sievenpiper_Journal,Patel_Grbic_Analytical_journal}.

\section{Two-Dimensional Transformations}

In transformation electromagnetics \cite{pendry}, fields are transformed
from an initial state to a desired one via a change in material
parameters based on a coordinate transformation. The transformed material
tensors ($\overline{\overline{\mu^{\prime\prime}}}$and $\overline{\overline{\epsilon^{\prime\prime\text{}}}}$)
are related to the initial material parameters ($\overline{\overline{\mu}}$
and $\overline{\overline{\epsilon}}$) in the following manner:
\begin{equation}
\overline{\overline{\mu^{\prime\prime}}}=\frac{\overline{\overline{J}}\ \overline{\overline{\mu}}(\overline{\overline{J}})^{T}}{|\overline{\overline{J}}|}\;\;\;\;\;\;\overline{\overline{\epsilon^{\prime\prime}}}=\frac{\overline{\overline{J}}\ \overline{\overline{\epsilon}}(\overline{\overline{J}})^{T}}{|\overline{\overline{J}}|},\label{eq:mu_eps_transform}
\end{equation}
where
\begin{equation}
\overline{\overline{J}}=\left(\begin{array}{ccc}
\frac{\partial x^{\prime\prime}}{\partial x} & \frac{\partial x^{\prime\prime}}{\partial y} & \frac{\partial x^{\prime\prime}}{\partial z}\\
\frac{\partial y^{\prime\prime}}{\partial x} & \frac{\partial y^{\prime\prime}}{\partial y} & \frac{\partial y^{\prime\prime}}{\partial z}\\
\frac{\partial z^{\prime\prime}}{\partial x} & \frac{\partial z^{\prime\prime}}{\partial y} & \frac{\partial z^{\prime\prime}}{\partial z}
\end{array}\right),
\end{equation}
is the Jacobian of the transformation from the $(x,y,z)$ coordinate
system to the $(x^{\prime\prime},y^{\prime\prime},z^{\prime\prime})$
system. When two-dimensional transformations are applied in the $x-y$
plane, the Jacobian reduces to
\begin{equation}
\overline{\overline{J}}=\left(\begin{array}{cc}
\frac{\partial x^{\prime\prime}}{\partial x} & \frac{\partial x^{\prime\prime}}{\partial y}\\
\frac{\partial y^{\prime\prime}}{\partial x} & \frac{\partial y^{\prime\prime}}{\partial y}
\end{array}\right)=\left(\begin{array}{cc}
J_{11} & J_{12}\\
J_{21} & J_{22}
\end{array}\right).
\end{equation}
In transformation electromagnetics, material parameters transform
as in \eqref{eq:mu_eps_transform}. However, when designing TISs,
the surface impedance/admittance is the quantity of interest rather
than material parameters. Therefore, we must find how the surface admittance
transforms. Transformation electromagnetics dictates that the transformed
fields are related to the initial fields as \cite{Werner_TO_overview,Kuprel}:

\begin{equation}
\overline{E}=\overline{\overline{J}}^{T}\overline{E}^{\prime\prime},\label{eq:E_field_Jacobian compact}
\end{equation}

\begin{equation}
\overline{H}=\overline{\overline{J}}^{T}\overline{H}^{\prime\prime}
\label{eq:H_field_Jacobian compact}
\end{equation}
or equivalently,

\begin{equation}
\left(\begin{array}{c}
E_{x}\\
E_{y}
\end{array}\right)=\left(\begin{array}{cc}
J_{11} & J_{21}\\
J_{12} & J_{22}
\end{array}\right)\left(\begin{array}{c}
E_{x}^{\prime\prime}\\
E_{y}^{\prime\prime}
\end{array}\right),\label{eq:E_field_Jacobian}
\end{equation}

\begin{equation}
\left(\begin{array}{c}
H_{x}\\
H_{y}
\end{array}\right)=\left(\begin{array}{cc}
J_{11} & J_{21}\\
J_{12} & J_{22}
\end{array}\right)\left(\begin{array}{c}
H_{x}^{\prime\prime}\\
H_{y}^{\prime\prime}
\end{array}\right).\label{eq:H_field_Jacobian}
\end{equation}
Rearranging the magnetic field components in \eqref{eq:H_field_Jacobian},
yields

\begin{equation}
\begin{split}\left(\begin{array}{c}
-H_{y}\\
H_{x}
\end{array}\right) & =\left(\begin{array}{cc}
J_{22} & -J_{12}\\
-J_{21} & J_{11}
\end{array}\right)\left(\begin{array}{c}
-H_{y}^{\prime\prime}\\
H_{x}^{\prime\prime}
\end{array}\right)\\
&=|\overline{\overline{J}}|\overline{\overline{J}}^{-1}\left(\begin{array}{c}
-H_{y}^{\prime\prime}\\
H_{x}^{\prime\prime}
\end{array}\right).
\end{split}
\label{eq:H_field_Jacob_properform}
\end{equation}
Substituting \eqref{eq:E_field_Jacobian} and \eqref{eq:H_field_Jacob_properform}
into the tensor admittance boundary condition (\ref{eq: Y boundary condition}) yields
\begin{equation}
\left(\begin{array}{c}
-H_{y}^{\prime\prime}\\
H_{x}^{\prime\prime}
\end{array}\right)=\overline{\overline{Y}}_{surf}^{\prime\prime}\left(\begin{array}{c}
E_{x}^{\prime\prime}\\
E_{y}^{\prime\prime}
\end{array}\right)=\frac{\overline{\overline{J}}\ \overline{\overline{Y}}_{surf}\overline{\overline{J}}^{T}}{|\overline{\overline{J}}|}\left(\begin{array}{c}
E_{x}^{\prime\prime}\\
E_{y}^{\prime\prime}
\end{array}\right).\label{eq:transformed Y}
\end{equation}
Comparing equations \eqref{eq: Y boundary condition} and \eqref{eq:transformed Y}
reveals that the transformation electromagnetics method
transforms the surface admittance in the same manner that $\overline{\overline{\epsilon}}$
and $\overline{\overline{\mu}}$ are transformed in \eqref{eq:mu_eps_transform}. That is,
\begin{equation}
\overline{\overline{Y}}_{surf}^{\prime\prime}=\frac{\overline{\overline{J}}\ \overline{\overline{Y}}_{surf}\overline{\overline{J}}^{T}}{|\overline{\overline{J}}|}.
\end{equation}

The transverse resonance equation that determines the guided modes,
for propagation along the $x$-axis of an idealized TIBC \cite{Patel_Grbic_Effective_journal} is given by
\begin{equation}
\left(\begin{array}{cc}
Y_{xx} & Y_{xy}\\
Y_{yx} & Y_{yy}
\end{array}\right)\left(\begin{array}{c}
E_{x}\\
E_{y}
\end{array}\right)=\left(\begin{array}{cc}
Y_{0}\frac{k_{0}}{k_{z}} & 0\\
0 & Y_{o}\frac{k_{z}}{k_{0}}
\end{array}\right)\left(\begin{array}{c}
E_{x}\\
E_{y}
\end{array}\right),\label{eq:transverse resonance}
\end{equation}
where $Y_{0}=\sqrt{\frac{\epsilon_{0}}{\mu_{0}}}$, and $k_{0}=k_{t}^{2}+k_{z}^{2}$ .
In general, the transverse wave number, $k_{t}$, is given by $k_{t}^{2}=k_{x}^{2}+k_{y}^{2}$ but in this particular case, $k_t = k_x$.
The matrix on the right-hand-side (RHS) of \eqref{eq:transverse resonance}
contains the $TM$ and $TE$ admittances of free space. %
{} Manipulating both sides of \eqref{eq:transverse resonance} yields
\begin{equation}
\begin{split} & \left[\frac{\overline{\overline{J}}}{|\overline{\overline{J}}|}\left(\begin{array}{cc}
Y_{xx} & Y_{xy}\\
Y_{yx} & Y_{yy}
\end{array}\right)\overline{\overline{J}}^{T}\right]\left(\left(\overline{\overline{J}}^{T}\right)^{-1}\left(\begin{array}{c}
E_{x}\\
E_{y}
\end{array}\right)\right)\\
 & =\frac{\overline{\overline{J}}}{|\overline{\overline{J}}|}\left(\begin{array}{cc}
Y_{o}\frac{k_{o}}{k_{z}} & 0\\
0 & Y_{o}\frac{k_{z}}{k_{o}}
\end{array}\right)\overline{\overline{J}}^{T}\left(\left(\overline{\overline{J}}^{T}\right)^{-1}\left(\begin{array}{c}
E_{x}\\
E_{y}
\end{array}\right)\right).
\end{split}
\label{eq:blah}
\end{equation}
The term in square brackets on the LHS of \eqref{eq:blah} can be
substituted with \eqref{eq:transformed Y}, yielding the following
equation,

\begin{equation}
\begin{split} & \left(\begin{array}{cc}
Y_{xx}^{\prime\prime} & Y_{xy}^{\prime\prime}\\
Y_{yx}^{\prime\prime} & Y_{yy}^{\prime\prime}
\end{array}\right)\left(\begin{array}{c}
E_{x}^{\prime\prime}\\
E_{y}^{\prime\prime}
\end{array}\right)\\
&=\left[\frac{\overline{\overline{J}}}{|\overline{\overline{J}}|}\left(\begin{array}{cc}
Y_{o}\frac{k_{o}}{k_{z}} & 0\\
0 & Y_{o}\frac{k_{z}}{k_{o}}
\end{array}\right)\overline{\overline{J}}^{T}\right]\left(\begin{array}{c}
E_{x}^{\prime\prime}\\
E_{y}^{\prime\prime}
\end{array}\right).
\end{split}
\label{eq:eq:transformfieldsabove}
\end{equation}
Therefore, not only is the surface admittance, $\overline{\overline{Y}}_{surf}$, transformed but so is the free space above the surface (term in square brackets of \eqref{eq:eq:transformfieldsabove}), to satisfy the guidance condition. This is impractical, since in many applications the space above the impedance surface is fixed: typically free space. This conclusion is verified through full-wave simulation
in Section IV. The transformation of free space above the surface is not needed for two-dimensional transformation electromagnetics devices based on transmission lines \cite{Gok_Tensor_TL,George_zedler_2D_TO,Kwon_Tensor_TL,liu_transmission_line_transformation,Selvan_sheared,Selvan_skewed}, since the fields are confined to the surface dimensions (i.e. $k_z =0$).

\begin{figure}
\begin{centering}
\includegraphics[width=3.4in]{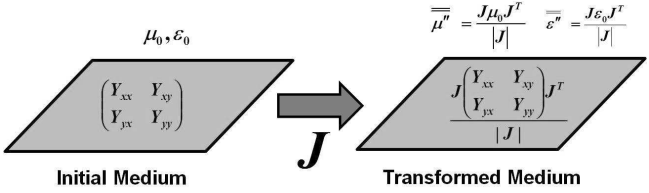}
\par\end{centering}
\caption{Transforming the surface and the space above it via the traditional transformation electromagnetics
method \eqref{eq:transformed Y}. An alternate method that does not transform the space above
the surface, but rather the TIBC alone is presented in Section III.}
\label{Fig:freespace_compression}
\end{figure}

\section{Transformation Electromagnetics Applied to an Idealized Tensor Impedance Boundary Condition
(TIBC)}

In the previous section, it was shown that the transformed surface
admittance ($\bar{\bar{Y}}_{surf}^{\prime\prime}$) can be found from
an initial surface impedance ($\bar{\bar{Y}}_{surf}$) in the same
manner that the transformed material parameters are computed. However, to maintain the guidance condition, the free space above the
surface must also be transformed. This section proposes an alternative design
approach. In this alternative approach, tensor impedance entries ($\eta_{xx}^{\prime\prime},\eta_{xy}^{\prime\prime}=\eta_{yy}^{\prime\prime},$
and $\eta_{yy}^{\prime\prime}$) are found that support the spatially
varying wave vector and Poynting vector of the transformation electromagnetics
device, while maintaining free space above the surface.

A plane wave's wave vector and Poynting vector tangential to the surface transform
as \cite{Gok_MTT_2013}:
\begin{equation}
\overline{k_{t}^{\prime\prime}}=\begin{pmatrix}k_{x}^{\prime\prime}\\
k_{y}^{\prime\prime}
\end{pmatrix}=(J^{T})^{-1}\overline{k_{t}},\label{eq:jacobian k-1}
\end{equation}
\begin{equation}
\overline{S_{t}^{\prime\prime}}=\begin{pmatrix}S_{x}^{\prime\prime}\\
S_{y}^{\prime\prime}
\end{pmatrix}=\left(\frac{J}{|J|}\right)\overline{S_{t}}.\label{eq:jacobian S-1}
\end{equation}
At a given spatial coordinate, the Poynting vector points at an angle,
$\theta_{power}^{\prime\prime}$, with respect to the x-axis,
\begin{equation}
\frac{S_{y}^{\prime\prime}}{S_{x}^{\prime\prime}}=\tan(\theta_{power}^{\prime\prime})=b.\label{eq:sy over sx equals b-1}
\end{equation}
Similarly, the transformed wave vector points at an angle, $\theta_{k_t}^{\prime\prime}=k_{y}^{\prime\prime}/k_{x}^{\prime\prime}$
with respect to the $x$-axis. 
 In addition to supporting the transformed wave vector and Poynting
vector, the tensor impedance entries ($\eta_{xx}^{\prime\prime},\eta_{xy}^{\prime\prime}=\eta_{yy}^{\prime\prime},$
and $\eta_{yy}^{\prime\prime}$) must also satisfy the guidance condition
for propagation along the surface.

\subsection{Propagation along TIBCs}

The following eigenvalue equation ((17) in \cite{Patel_Grbic_Analytical_journal}) can be written to find the modes
supported by a TIBC:
\begin{equation}
\begin{pmatrix}b_{11} & b_{12}\\
b_{21} & b_{22}
\end{pmatrix}\begin{pmatrix}E_{z}^{\prime\prime}\\
H_{z}^{\prime\prime}
\end{pmatrix}=0,\label{eq:eigenval}
\end{equation}
where
\begin{equation}
\begin{split}b_{11} & =k_{x}^{\prime\prime}k_{z}^{\prime\prime}+\frac{k_{x}^{\prime\prime}k_{0}\eta_{xx}^{\prime\prime}}{\eta_{0}}+\frac{k_{y}^{\prime\prime}k_{0}\eta_{xy}^{\prime\prime}}{\eta_{0}}\\
b_{12} & =k_{0}k_{y}^{\prime\prime}\eta_{0}+k_{y}^{\prime\prime}k_{z}^{\prime\prime}\eta_{xx}^{\prime\prime}-k_{x}^{\prime\prime}k_{z}^{\prime\prime}\eta_{xy}^{\prime\prime}\\
b_{21} & =k_{y}^{\prime\prime}k_{z}^{\prime\prime}+\frac{k_{x}^{\prime\prime}k_{0}\eta_{yx}^{\prime\prime}}{\eta_{0}}+\frac{k_{y}^{\prime\prime}k_{0}\eta_{yy}^{\prime\prime}}{\eta_{0}}\\
b_{22} & =-k_{0}k_{x}^{\prime\prime}\eta_{0}+k_{y}^{\prime\prime}k_{z}^{\prime\prime}\eta_{yx}^{\prime\prime}-k_{x}^{\prime\prime}k_{z}^{\prime\prime}\eta_{yy}^{\prime\prime}.
\end{split}
\end{equation}
The eigenvalue equation above is found by expressing the tangential field components
($E_{x}^{\prime\prime},E_{y}^{\prime\prime},H_{x}^{\prime\prime},$
and $H_{y}^{\prime\prime}$) in terms of the normal field components ($E_{z}^{\prime\prime}$
and $H_{z}^{\prime\prime}$) corresponding to $TM$ and $TE$ fields,
respectively \cite{Patel_Grbic_Analytical_journal,Kong}. It should be noted that the double primes now denote field quantities corresponding to the transformed wave vector \eqref{eq:jacobian k-1} and Poynting vector \eqref{eq:jacobian S-1}, not the transformed fields (given by \eqref{eq:E_field_Jacobian compact} and \eqref{eq:H_field_Jacobian compact}) from transformation electromagnetics. From \eqref{eq:eigenval}, the dispersion equation of a TIBC can be derived \cite{Bilow,Patel_Grbic_Analytical_journal}:
\begin{equation}
b_{11}b_{22}-b_{12}b_{21}=0.\label{eq:TIS dispersion}
\end{equation}
The group velocity along a TIBC can be found by differentiating the dispersion equation \eqref{eq:TIS dispersion} to find
\begin{equation}
\overline{v_g}=v_{g_x} \hat{x} +v_{g_y} \hat{y}=\frac{\partial \omega}{\partial k_x} \hat{x} +\frac{\partial \omega}{\partial k_y} \hat{y} .\label{eq:TIS groupvelocity}
\end{equation}
The direction of power flow along a TIBC can then be expressed as \cite{Patel_Grbic_Power_Flow}:
\begin{equation}
\begin{split} & \frac{v_{g_y} }{v_{g_x} }= \tan\theta_{power}^{\prime\prime}=\\
 & \frac{k_{z}^{\prime\prime}Y_{0}(k_{x}^{\prime\prime}(Y_{xy}^{\prime\prime}+Y_{yx}^{\prime\prime})+2k_{y}^{\prime\prime}Y_{yy}^{\prime\prime})+k_{0}k_{y}^{\prime\prime}(Y_{0}^{2}+\det Y_{surf}^{\prime\prime})}{k_{z}^{\prime\prime}Y_{0}(k_{y}^{\prime\prime}(Y_{xy}^{\prime\prime}+Y_{yx}^{\prime\prime})+2k_{x}^{\prime\prime}Y_{xx}^{\prime\prime})+k_{0}k_{x}^{\prime\prime}(Y_{0}^{2}+\det Y_{surf}^{\prime\prime})},
\end{split}
\label{eq:SS:direction power_primes}
\end{equation}
The eigenvalue equation \eqref{eq:eigenval} will be
used to design TIBCs that support surface waves with the transformed
wave vector and Poynting vector distributions given by \eqref{eq:jacobian k-1}
and \eqref{eq:jacobian S-1}.

\subsection{Design Approach}

In transformation electromagnetics, the transformed material parameters
are derived from an initial medium. This initial medium is typically
free space. Since the intent here is to apply transformation electromagnetics
to TIBCs, an initial isotropic surface impedance,
\begin{equation}
\overline{\overline{\eta}}_{surf}=\begin{pmatrix}\eta & 0\\
0 & \eta
\end{pmatrix},\label{eq:iso imp}
\end{equation}
in free space is chosen that supports a surface wave at a desired frequency
of operation. The surface impedance supporting a $TM$ surface
wave is given by
\begin{equation}
\eta=\eta_{0}\sqrt{1-\left(\frac{k_{t}}{k_{0}}\right)^{2}}.\label{eq: surface impedance from kt-1}
\end{equation}
The tangential wave number ($k_{t}$) along the surface is chosen
to be greater than that of free space $(k_{t}>k_{0})$ to ensure a
bound surface wave. Next, a surface impedance,
\begin{equation}
\overline{\overline{\eta}}_{surf}^{\prime\prime}=\begin{pmatrix}\eta_{xx}^{\prime\prime} & \eta_{xy}^{\prime\prime}\\
\eta_{yx}^{\prime\prime} & \eta_{yy}^{\prime\prime}
\end{pmatrix},
\end{equation}
is found which supports the transformed wave vector and Poynting vector distributions on the surface. By writing the Poynting vector
components $(S_{x}^{\prime\prime},S_{y}^{\prime\prime})$ in terms
of $E_{z}^{\prime\prime}$ and $H_{z}^{\prime\prime}$, \eqref{eq:sy over sx equals b-1}
can be recast as
\begin{equation}
\begin{split}\frac{E{}_{z}^{\prime\prime}}{H_{z}^{\prime\prime}} & =-\frac{\eta_{0}}{k_{0}(bk_{x}^{\prime\prime}-k_{y}^{\prime\prime})}\left[(k_{x}^{\prime\prime}+bk_{y}^{\prime\prime})k_{z}^{\prime\prime}\right.\\
&\left.\pm j\sqrt{-k_{0}^{2}(-bk_{x}^{\prime\prime}+k_{y}^{\prime\prime})^{2}-(k_{x}^{\prime\prime}+bk_{y}^{\prime\prime})^{2}k_{z}^{\prime\prime2}}\right].
\end{split}
\label{eq:EzoverHzratio}
\end{equation}
Therefore, the transformed wave vector $(k_{x}^{\prime\prime},k_{y}^{\prime\prime})$
and direction of the Poynting vector ($b=\tan(\theta_{power}^{\prime\prime}$))
along the surface, uniquely define the ratio
of the normal electric to magnetic fields (ratio of $TM$ to $TE$ fields) supported by the TIBC. Even
though the isotropic surface impedance supports a $TM$ wave only, the anisotropic
surface impedance can support a mixture of $TM$ and $TE$ waves, as indicated by
\eqref{eq:EzoverHzratio}. %
Equation \eqref{eq:EzoverHzratio} first appears in \cite{Patel_IMS2013_transformation} but there, it contains a typographical error.
Substituting \eqref{eq:EzoverHzratio} into \eqref{eq:eigenval}
yields two (out of three) equations for finding the surface impedance entries: $\eta_{xx}^{\prime\prime},\eta_{xy}^{\prime\prime}=\eta_{yx}^{\prime\prime},$
and $\eta_{yy}^{\prime\prime}$. Setting the determinant of the transformed
surface impedance tensor equal to the square of the initial surface
impedance ($\eta$), results in a third equation,
\begin{equation}
\eta_{xx}^{\prime\prime}\eta_{yy}^{\prime\prime}-\eta_{xy}^{\prime\prime}\eta_{yx}^{\prime\prime}=\eta^{2}.\label{eq:TO condition}
\end{equation}
The transformed surface impedance entries can now be found using these
three equations. This condition on the determinant of the surface
impedance is analogous to the condition on the permittivity and permeability
tensors in transformation electromagnetics devices \cite{Gok_MTT_2013}.
Solving this system of three equations yields the surface impedance
tensor necessary (at each point on the surface) to ensure the desired
distributions of wave vector \eqref{eq:jacobian k-1} and direction
of power flow \eqref{eq:jacobian S-1} along the surface. Alternatively,
the system of three equations needed to find the surface impedance
entries can be chosen as: the dispersion equation \eqref{eq:TIS dispersion},
the direction of power flow \eqref{eq:SS:direction power_primes}, and \eqref{eq:TO condition}.

\section{Example: A Beam-shifting Surface using a TIBC}

\begin{figure}
\begin{centering}
\includegraphics[width=3.0in]{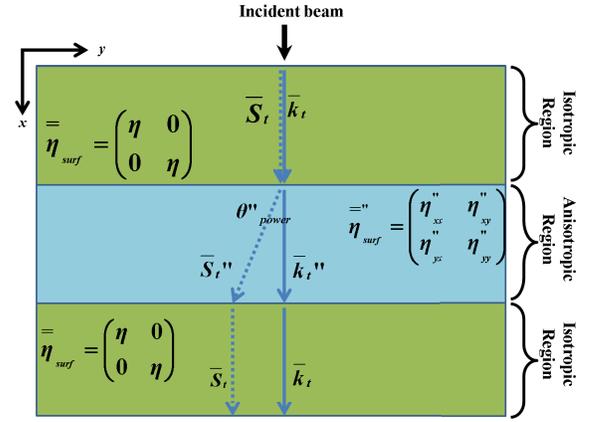}
\par\end{centering}
\caption{A beam-shifting surface consisting of three regions. Two different
surfaces need to be designed; one isotropic and one anisotropic.
The anisotropic surface is designed to bend the incident beam by $\theta_{power}^{\prime\prime}$.}
\label{Fig: beamshifter diagram}
\end{figure}

In this section, a transformation-based beam-shifter \cite{Gok_beam_shifter_Journal,Kong_beamshifter} is designed
using TIBCs. The device can bend a surface-wave beam by an angle of
$\theta_{power}^{\prime\prime}$. The device consists of three regions
(as shown in Fig. \ref{Fig: beamshifter diagram}): an anisotropic region
with surface impedance $\bar{\bar{\eta}}_{surf}^{\prime\prime}$ sandwiched
between two isotropic regions with surface impedance $\bar{\bar{\eta}}_{surf}$.
In the uppermost isotropic region, propagation is set to be purely
in the $x$-direction. The wave number is chosen to be $k_{x}=1.1882 k_0=248.85$
rad/m at $10$ GHz, to ensure a tightly bound wave. The corresponding
surface impedance is given by \eqref{eq: surface impedance from kt-1}:
\begin{equation}
\eta_{surf}=j\begin{pmatrix}241.91 & 0\\
0 & 241.91
\end{pmatrix}\Omega.\label{eq:isotropic region surface impedance}
\end{equation}
This surface will support a $TM$ surface wave with the following
propagation characteristics:
\begin{equation}
\overline{k_{t}}=\begin{pmatrix}k_{x}\\
k_{y}
\end{pmatrix}=\begin{pmatrix}248.85\\
0
\end{pmatrix}rad/m,\label{eq:isotropic_phase_delay}
\end{equation}
\begin{equation}
\overline{S_{t}}=\begin{pmatrix}S_{x}\\
S_{y}
\end{pmatrix}=\begin{pmatrix}S_{x}\\
0
\end{pmatrix}W/m^{2}.\label{eq:isotropic_poynting}
\end{equation}

\begin{figure}
\begin{centering}
\includegraphics[width=3in]{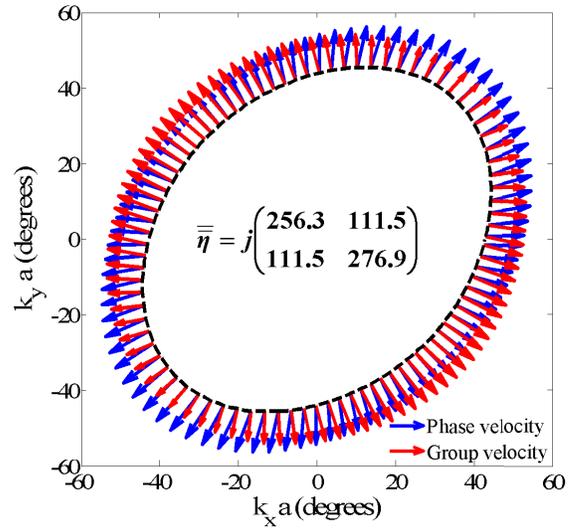}
\par\end{centering}
\caption{The isofrequency dispersion contour at 10 GHz for the idealized inductive TIBC (anisotropic
surface) corresponding to the designed TIBC beam-shifter (\ref{eq:aniso tensor}).
Arrows point in the directions of group velocity (red) and phase velocity (blue).
The group and phase velocities co-align along the principal axes of
the surface. The length of the red arrows represent the normalized
magnitude of the group velocity. For propagation along the $x$-axis
($\theta_{k_t}^{\prime\prime}=0$), the angle between the group velocity vector and the phase velocity vector is $-13.93^{\circ}$ as designed.}
\label{Fig:TIS beamshifter_dispersion}
\end{figure}

The anisotropic region is designed by finding the anisotropic surface
impedance tensor ($\bar{\bar{\eta}}_{surf}^{\prime\prime}$) needed
to bend the beam by an angle $\theta_{power}^{\prime\prime}$ (Fig.
\ref{Fig: beamshifter diagram}). A coordinate transformation is applied
to $\bar{k_{t}}$ and $\bar{S_{t}}$ to find the transformed tangential
wave vector ($\bar{k_{t}^{\prime\prime}}$) and Poynting vector ($\bar{S_{t}^{\prime\prime}}$)
in the anisotropic region. The Jacobian of the coordinate transformation
governing the anisotropic region of the beam-shifting device is given
by \cite{Werner_TO_overview}
\begin{equation}
J=\begin{pmatrix}1 & 0\\
b & 1
\end{pmatrix},\label{eq:jacobian}
\end{equation}
where $b=\tan(\theta_{power}^{\prime\prime}).$ The beam-shift angle
is chosen to be $\theta_{power}=-13.93^{\circ}$or equivalently, $b=-0.248$.
Applying the transformation to $\overline{k}_{t}$ and $\overline{S_{t}}$,
using \eqref{eq:jacobian k-1} and \eqref{eq:jacobian S-1} yields,
\begin{equation}
\overline{k_{t}^{\prime\prime}}=\begin{pmatrix}k_{x}^{\prime\prime}\\
k_{y}^{\prime\prime}
\end{pmatrix}=\begin{pmatrix}248.85\\
0
\end{pmatrix}\label{eq:transformed k}
\end{equation}
and
\begin{equation}
\overline{S_{t}^{\prime\prime}}=\begin{pmatrix}S_{x}^{\prime\prime}\\
S_{y}^{\prime\prime}
\end{pmatrix}=\begin{pmatrix}S_{x}\\
bS_{x}
\end{pmatrix}.\label{eq:Poynting beam shift}
\end{equation}
Applying the design procedure described in the previous section yields
the following surface impedance tensor for the anisotropic region:
\begin{equation}
\eta_{surf}^{\prime\prime}=\begin{pmatrix}\eta_{xx}^{\prime\prime} & \eta_{xy}^{\prime\prime}\\
\eta_{yx}^{\prime\prime} & \eta_{yy}^{\prime\prime}
\end{pmatrix}=j\begin{pmatrix}256.3 & 111.5\\
111.5 & 276.9
\end{pmatrix}\Omega.\label{eq:aniso tensor}
\end{equation}
The dispersion contour for the anisotropic region is shown in Fig. \ref{Fig:TIS beamshifter_dispersion}.

\begin{figure}
\begin{centering}
\includegraphics[width=3in]{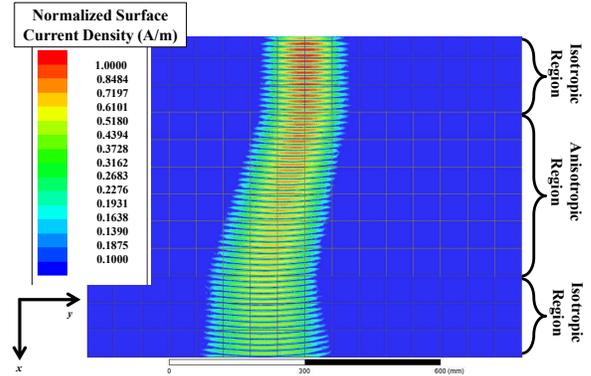}
\par\end{centering}
\caption{Normalized surface current density for the beam-shifting surface.
The incoming beam is deflected by $-13.93^{\circ}$ in the anisotropic
region. The total size of the surface is $96\times72$ cm ($32\lambda_0\times24\lambda_0$). Each isotropic region
is $96\times18$ cm ($32\lambda_0\times6\lambda_0$). The dimensions of the anisotropic region are $96\times36$ cm ($32\lambda_0\times12\lambda_0$). }
\label{Fig: beamshift sim results}
\end{figure}

The beam-shifter was simulated using Ansys HFSS. The isotropic and
anisotropic regions (as shown in Fig. \ref{Fig: beamshifter diagram})
were modeled using the screening impedance boundary.
The boundaries of the simulation domain were terminated with radiation
boundaries, and one edge was illuminated with a Gaussian beam. The
results of the simulation at 10 GHz are shown in Fig. \ref{Fig: beamshift sim results}.
As expected, the Gaussian excitation couples energy into the uppermost
isotropic surface, and a surface wave propagates in the $x$-direction.
Upon encountering the anisotropic region, the beam is refracted by
$-13.93^{\circ}$. To an observer at the far edge of the lower isotropic
region, (edge opposite the source), the source appears to have
shifted laterally.

Had the surface admittance \eqref{eq:isotropic region surface impedance}
been transformed by \eqref{eq:transformed Y}, the transformed surface
impedance would be:
\begin{equation}
\eta_{surf}^{\prime\prime}=\begin{pmatrix}\eta_{xx}^{\prime\prime} & \eta_{xy}^{\prime\prime}\\
\eta_{yx}^{\prime\prime} & \eta_{yy}^{\prime\prime}
\end{pmatrix}=j\begin{pmatrix}256.79 & 59.99\\
59.99 & 241.91
\end{pmatrix}\Omega.\label{eq:aniso tensor_transform_Free_space}
\end{equation}
This surface impedance tensor does not satisfy the guidance condition
at 10 GHz unless the free space above the surface is transformed to
\begin{equation}
\epsilon^{\prime\prime}=\epsilon_{0}\begin{pmatrix}1 & b\\
b & b^{2}+1
\end{pmatrix}=\epsilon_{0}\begin{pmatrix}1 & -0.248\\
-0.248 & 1.062
\end{pmatrix}\label{eq:epsilon_stretched}
\end{equation}
\begin{equation}
\mu^{\prime\prime}=\mu_{0}\begin{pmatrix}1 & b\\
b & b^{2}+1
\end{pmatrix}=\mu_{0}\begin{pmatrix}1 & -0.248\\
-0.248 & 1.062
\end{pmatrix}\label{eq:mu stretched}
\end{equation}
via \eqref{eq:mu_eps_transform}.
This fact is verified using HFSS's eigenmode solver. A unit cell of the TIBC given by \eqref{eq:aniso tensor_transform_Free_space} is implemented in HFSS with a screening impedance and the medium above the surface is assigned the anisotropic material parameters described by \eqref{eq:epsilon_stretched} and \eqref{eq:mu stretched}.
When the phase delay corresponding to \eqref{eq:transformed k} is stipulated along the $x-$direction of the surface, an eigenfrequency of 10 GHz was found by the eigenmode solver. This verifies that for a surface transformed using \eqref{eq:transformed Y}, the guidance condition is only satisfied when the free space above the surface is also transformed. Additionally, finding the ratio of $S_{y}^{\prime\prime}$
to $S_{x}^{\prime\prime}$ from the simulation verifies the direction
of power flow as $\theta_{power}^{\prime\prime}=-13.93^{\circ}.$ When
the free space above the surface is left untransformed in simulation,
the guidance condition is satisfied at 9.874 GHz, which agrees with
analytical predictions from the dispersion equation. At this frequency, the direction of power flow is $\theta_{power}^{\prime\prime}=-8.37^{\circ}.$

In the next section, a beam-shifter is implemented with a PCTIS. In the case of a PCTIS, the unknowns
are the sheet admittance tensor entries ($Y_{xx}^{s\prime\prime}$, $Y_{xy}^{s\prime\prime}$,
and $Y_{yy}^{s\prime\prime}$) rather than the surface admittance tensor entries.

\section{Transformation Electromagnetics Applied to Printed-Circuit Tensor
Impedance Surfaces (PCTISs)}

\begin{figure}
\begin{centering}
\includegraphics[width=3in]{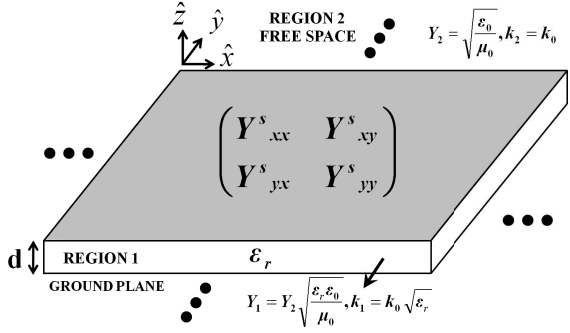}
\par\end{centering}
\caption{PCTIS consisting of a tensor
sheet impedance over a grounded dielectric substrate. The tensor sheet
impedance/admittance, which models a generalized metallic cladding,
is denoted with a superscript `s'.}
\label{Fig:PCTIS model}
\end{figure}

In this section, a procedure for designing transformation electromagnetics
devices using PCTISs is presented. A PCTIS consists of a tensor impedance
sheet over a grounded dielectric substrate, where the tensor sheet
impedance models a patterned metallic cladding. As shown in the analytical
model of a PCTIS (see Fig. \ref{Fig:PCTIS model}), the quantities of
interest are the sheet admittance entries. The effective surface admittance
of a PCTIS was related to the surface admittance of a TIBC
in \cite{Patel_Grbic_Effective_journal}. It was found that a PCTIS
exhibits spatial dispersion due to its electrical thickness. As a
result of this spatial dispersion, a PCTIS can have the same surface
impedance as a TIBC, but a different direction of power flow \cite{Patel_Grbic_Power_Flow}.
The design method presented in the section is analogous to the design
procedure for TIBCs from Section III. However, in the case of a PCTIS,
one must find the sheet admittance entries ($Y_{xx}^{s\prime\prime}$, $Y_{xy}^{s\prime\prime}=Y_{yx}^{s\prime\prime}$,
and $Y_{yy}^{s\prime\prime}$) that support the transformed wave vector
and Poynting vector distributions of a transformation electromagnetics
device.

\subsection{Propagation along PCTISs}

The modes supported by a PCTIS can be found from eigenvalue equation (20) in \cite{Patel_Grbic_Analytical_journal}. The dispersion
equation for a PCTIS can be derived from this eigenvalue equation as \cite{Patel_Grbic_Analytical_journal,Patel_Grbic_Effective_journal},
\begin{equation}
\begin{split} & 4\epsilon_{1}k_{1}^{2}k_{2}^{2}k_{z1}^{\prime\prime}k_{z2}^{\prime\prime}\mu_{2}\w\cos^{2}(k_{z1}^{\prime\prime}d)\\
 & +j\sin(2k_{z1}^{\prime\prime}d)\\
 & \quad\quad[2\epsilon_{1}k_{2}^{4}(k_{z1}^{\prime\prime})^{2}\mu_{1}\w+2\epsilon_{2}k_{1}^{4}(k_{z2}^{\prime\prime})\mu_{2}\w\\
 & \quad\quad+k_{1}^{4}k_{2}^{2}k_{z2}^{\prime\prime}\mu_{2}Y_{xx}^{s\prime\prime}+\epsilon_{1}k_{2}^{2}(k_{z1}^{\prime\prime})^{2}k_{z2}^{\prime\prime}\mu_{1}\mu_{2}\w^{2}Y_{xx}^{s\prime\prime}\\
 & \quad\quad+k_{1}^{4}k_{2}^{2}k_{z2}^{\prime\prime}\mu_{2}Y_{yy}^{s\prime\prime}+\epsilon_{1}k_{2}^{2}(k_{z1}^{\prime\prime})^{2}k_{z2}^{\prime\prime}\mu_{1}\mu_{2}\w^{2}Y_{yy}^{s\prime\prime}\\
 & \quad\quad-k_{2}^{2}k_{z2}^{\prime\prime}\mu_{2}(k_{1}^{4}-\epsilon_{1}(k_{z1}^{\prime\prime})^{2}\mu_{1}\w^{2})(Y_{xx}^{s\prime\prime}-Y_{yy}^{s\prime\prime})\cos(2\theta_{k}^{\prime\prime})\\
 & \quad\quad-k_{2}^{2}k_{z2}^{\prime\prime}\mu_{2}(k_{1}^{4}-\epsilon_{1}(k_{z1}^{\prime\prime})^{2}\mu_{1}\w^{2})(Y_{xy}^{s\prime\prime}+Y_{yx}^{s\prime\prime})\sin(2\theta_{k}^{\prime\prime})]\\
 & -2k_{1}^{2}k_{z}^{\prime\prime}\mu_{1}\sin^{2}(k_{z1}^{\prime\prime}d)\\
 & \quad\quad[2\epsilon_{2}k_{2}^{2}k_{z2}^{\prime\prime}\w+k_{2}^{4}Y_{xx}^{s\prime\prime}+\epsilon_{2}k_{z2}^{2}\mu_{2}\w^{2}Y_{xx}^{s\prime\prime}\\
 & \quad\quad-2k_{2}^{2}k_{z2}^{\prime\prime}\mu_{2}\w Y_{xy}^{s\prime\prime}Y_{yx}^{s\prime\prime}+k_{2}^{4}Y_{yy}^{s\prime\prime}+\epsilon_{2}(k_{z2}^{\prime\prime})^{2}\mu_{2}\w^{2}Y_{yy}^{s\prime\prime}\\
 & \quad\quad+2k_{2}^{2}k_{z2}^{\prime\prime}\mu_{2}\w Y_{xx}^{s\prime\prime}Y_{yy}^{s\prime\prime}\\
 & \quad\quad-(k_{2}^{4}-\epsilon_{2}(k_{z2}^{\prime\prime})^{2}\mu_{2}\w^{2})(Y_{xx}^{s\prime\prime}-Y_{yy}^{s\prime\prime})\cos(2\theta_{k}^{\prime\prime})\\
 & \quad\quad-(k_{2}^{4}-\epsilon_{2}(k_{z2}^{\prime\prime})^{2}\mu_{2}\w^{2})(Y_{xy}^{s\prime\prime}+Y_{yx}^{s\prime\prime})\sin(2\theta_{k}^{\prime\prime})]=0.
\end{split}
\label{eq:dispersion_whole_compact}
\end{equation}
Furthermore, the group velocity along a PCTIS was derived in \cite{Patel_Grbic_Power_Flow} by differentiating the dispersion equation with respect to $k_x$ and $k_y$. The $x$ and $y$ components of the group velocity along a PCTIS can be expressed compactly as:
\begin{equation}
\begin{split}
v_{gq}^{\text{PCTIS}}&=\frac{\partial\omega}{\partial k_{q}}\\
&=\frac{-B_7 +\frac{k_q}{k_{z2}}B_4-\xi_{1}B_1+\xi_{2}(B_2+B_6)-\xi_{4}B_3}{\chi_{1}B_1-\chi_{2}B_2-\chi_{3}B_6+\chi_{4}B_3+\nu B_4+\zeta B_5}
\end{split}
\label{eq:PCTIS:vgx}
\end{equation}
where
\begin{equation}
q = x \text{ or } y,
r=\left\{ \begin{array}{c}
y \text{ if } q=x\\
x\text{ if } q=y
\end{array}\right.,
\label{eq:AppA q r index}
\end{equation}
and $B_n$, $\chi$, $\zeta$, $\nu$ and $\xi$ terms are given in Table III in Appendix B of \cite{Patel_Grbic_Power_Flow}. The direction of power flow along a PCTIS is then given by
\begin{equation}
\begin{split}
\tan\theta_{s}^{\text{PCTIS}}=\frac{v_{gy}^{\text{PCTIS}}}{v_{gx}^{\text{PCTIS}}}.
\end{split}
\label{eq:PCTIS:direction power}
\end{equation}
The dispersion equation \eqref{eq:dispersion_whole_compact} and the equation above \eqref{eq:PCTIS:direction power} constitute two of the three equations needed to design a PCTIS that supports a surface wave with transformed wave vector and Poynting vector distributions. The third equation will be presented in the next subsection of this paper.

\subsection{Design Approach}

Similar to the design approach for TIBCs outlined in Section III, an
initial anisotropic medium must be stipulated. An isotropic sheet
admittance,

\begin{equation}
\overline{\overline{Y}}_{sheet}=\begin{pmatrix}Y^{s} & 0\\
0 & Y^{s}
\end{pmatrix},\label{eq:iso admttance pctis}
\end{equation}
is chosen to support a surface wave, with a desired transverse wave
number, $k_{t}$, at a chosen frequency of operation. For a $TM$ surface
wave, the isotropic sheet impedance is found from the $TM$ transverse
resonance equation \cite{Patel_Grbic_Effective_journal}

\begin{equation}
Y^{s}=Y_{surf}^{TM}+jY_{1}\frac{k_{1}}{k_{z1}}\cot(k_{z1}d).\label{eq: surface impedance from kt-1-1-1}
\end{equation}
or equivalently,

\begin{equation}
Y^{s}=\frac{Y_{0}}{\sqrt{1-\left(\frac{k_{t}}{k_{0}}\right)^{2}}}+jY_{1}\frac{k_{1}}{\sqrt{k_{1}^{2}-k_{t}^{2}}}\cot(\sqrt{k_{1}^{2}-k_{t}^{2}}d).\label{eq: surface impedance from kt-1-1}
\end{equation}
The transformed wave vector and Poynting vector are found from \eqref{eq:jacobian k-1}
and \eqref{eq:jacobian S-1}, respectively. Next, the sheet impedance
tensor,

\begin{equation}
\overline{\overline{Y}}_{sheet}^{\prime\prime}=\begin{pmatrix}Y_{xx}^{s\prime\prime} & Y_{xy}^{s\prime\prime}\\
Y_{yx}^{s\prime\prime} & Y_{yy}^{s\prime\prime}
\end{pmatrix},
\end{equation}
that supports the transformed wave vector and Poynting vector, is
found by solving a system of three equations: the dispersion equation
for a PCTIS \eqref{eq:dispersion_whole_compact}, the expression for the direction of
power flow along a PCTIS \eqref{eq:PCTIS:direction power},
and a condition on the determinant of the transformed sheet admittance \cite{Gok_MTT_2013},
\begin{equation}
Y_{xx}^{s\prime\prime}Y_{yy}^{s\prime\prime}-Y_{xy}^{s\prime\prime}Y_{yx}^{s\prime\prime}=(Y^{s})^{2}.\label{eq:TO condition PCTIS}
\end{equation}

Additionally, we must ensure that only a single mode is supported
by the PCTIS. That is, the higher order $TE$ mode should not be excited. The $TE$ mode cutoff occurs when $Y_{surf}^{TE}=0$. At cutoff, the $TE$ transverse resonance equation becomes

\begin{equation}
Y_{critical}^{s}=jY_{0}(\sqrt{\epsilon_r-1})\cot(k_{0}(\sqrt{\epsilon_r-1})d),\label{eq:critical-1}
\end{equation}
where $Y_{critical}^{s}$ is the sheet admittance at cutoff. The
eigenvalues ($Y_{\lambda_{1}}^{s}$ and $Y_{\lambda_{2}}^{s}$) of
$Y_{sheet}^{\prime\prime}$ can be found by diagonalizing $Y_{sheet}^{\prime\prime}$,

\begin{equation}
\overline{\overline{P}}^{-1}\overline{\overline{Y}}_{sheet}^{\prime\prime}\overline{\overline{P}}=\begin{pmatrix}Y_{\lambda_{1}}^{s} & 0\\
0 & Y_{\lambda_{2}}^{s}
\end{pmatrix},
\end{equation}
where $\overline{\overline{P}}$ is a matrix containing the eigenvectors
of $Y_{sheet}^{\prime\prime}$. In order to ensure that the higher order $TE$ mode is not excited, the eigenvalues ($Y_{\lambda_{1}}^{s}$ and $Y_{\lambda_{2}}^{s}$)
of $Y_{sheet}^{\prime\prime}$ must not exceed $Y_{critical}^{s}$.
When either of the eigenvalues is equal to $Y_{critical}^{s}$, the $TE$ mode can be excited. Beyond this resonance,
the surface impedance is capacitive and a $TE$ mode is supported in
addition to the $TM$ mode. In other words, the following conditions
must be satisfied in order to guarantee only one $TM$ mode exists:

\begin{equation}
Y_{\lambda_{1}}^{s}<Y_{critical}^{s}\label{eq:cond1}
\end{equation}
and
\begin{equation}
Y_{\lambda_{2}}^{s}<Y_{critical}^{s}\label{eq:cond2}
\end{equation}
where

\begin{equation}
Y_{\lambda_{1}}^{s}=\frac{Y_{xx}^{s\prime\prime}+Y_{yy}^{s\prime\prime}-\sqrt{{Y_{xx}^{s\prime\prime}}^{2}+4Y_{xy}^{s\prime\prime}Y_{yx}^{s\prime\prime}-2Y_{xx}^{s\prime\prime}Y_{yy}^{s\prime\prime}+{Y_{yy}^{s\prime\prime}}^{2}}}{2},
\end{equation}

\begin{equation}
Y_{\lambda_{2}}^{s}=\frac{Y_{xx}^{s\prime\prime}+Y_{yy}^{s\prime\prime}+\sqrt{{Y_{xx}^{s\prime\prime}}^{2}+4Y_{xy}^{s\prime\prime}Y_{yx}^{s\prime\prime}-2Y_{xx}^{s\prime\prime}Y_{yy}^{s\prime\prime}+{Y_{yy}^{s\prime\prime}}^{2}}}{2}.
\end{equation}
Simultaneously solving the three aforementioned equations under the
constraints of \eqref{eq:cond1} and \eqref{eq:cond2}, the tensor
sheet admittance entries ($Y_{xx}^{s\prime\prime}$, $Y_{xy}^{s\prime\prime}=Y_{yx}^{s\prime\prime}$,
and $Y_{yy}^{s\prime\prime}$) can be found. The choice of the isotropic
sheet ($Y^{s}$) may have to be adjusted in order to satisfy \eqref{eq:cond1} and \eqref{eq:cond2} in addition to the three equations. Essentially, this condition places a limitation on the beam-shift angles achievable
for a substrate with a given thickness and dielectric constant.

\section{Example: A Beam-shifter using a PCTIS}

\begin{figure}
\begin{centering}
\includegraphics[width=3in]{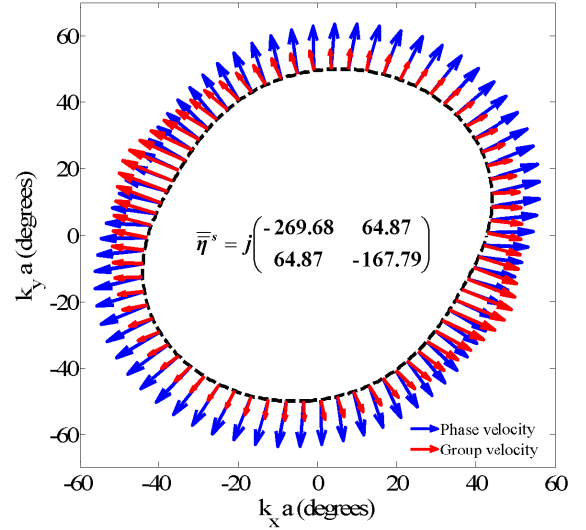}
\par\end{centering}
\caption{The isofrequency dispersion contour at 10 GHz for the anisotropic surface of the PCTIS beam-shifter (\ref{eq:aniso tensor}). Arrows point in the group velocity (red) and phase velocity (blue) directions. The group and
phase velocities co-align along the principal axes of the surface.
The length of the red arrows represent the normalized magnitude of
the group velocity. For propagation along the $x$-axis ($\theta_{k_t}^{\prime\prime}=0$),
the angle between the group velocity vector and the phase velocity vector is $-13.93^{\circ}$ as designed.}
\label{Fig:PCTIS beamshifter_dispersoin}
\end{figure}

In this section, a beam-shifter is designed using a PCTIS. The device
can bend a surface-wave beam by $-13.93^{\circ}$ at 10 GHz. It consists
of three regions, as shown in Fig 3. The PCTIS beam-shifter consists
of an isotropic sheet impedance in the upper and lower regions and
an anisotropic sheet impedance in the middle. The sheets are on a
1.27 mm thick grounded dielectric substrate with $\epsilon_{r}$=10.2.
In the isotropic region, propagation is chosen to be in the $x$-direction
with a transverse wave number of $k_x = 1.1882 k_0=248.85$ rad/m. The isotropic sheet
impedance, calculated using \eqref{eq:iso admttance pctis} and \eqref{eq: surface impedance from kt-1-1}
is

\begin{equation}
\eta_{sheet=}Y_{sheet}^{-1}=j\begin{pmatrix}-202.57 & 0\\
0 & -202.57
\end{pmatrix}\Omega.\label{eq:isotropic region PCTIS}
\end{equation}
The transformed wave and Poynting vectors are found to be

\begin{equation}
\overline{k_{t}^{\prime\prime}}=\begin{pmatrix}k_{x}^{\prime\prime}\\
k_{y}^{\prime\prime}
\end{pmatrix}=\begin{pmatrix}248.85\\
0
\end{pmatrix},
\end{equation}
and
\begin{equation}
\overline{S_{t}^{\prime\prime}}=\begin{pmatrix}S_{x}^{\prime\prime}\\
S_{y}^{\prime\prime}
\end{pmatrix}=\begin{pmatrix}S_{x}\\
bS_{x}
\end{pmatrix},\label{eq:Poynting beam shift-1 PCTIS}
\end{equation}
where $b=-0.248$.
Solving the system of three equations (\eqref{eq:dispersion_whole_compact}, \eqref{eq:PCTIS:direction power}, and \eqref{eq:TO condition PCTIS}) discussed in the design procedure
yields the following sheet impedance tensor for the anisotropic region:
\begin{equation}
\eta_{sheet=}^{\prime\prime}(Y_{sheet}^{\prime\prime})^{-1}=j\begin{pmatrix}-269.68 & 64.87\\
64.87 & -167.79
\end{pmatrix}\Omega.\label{eq:anisotropic region PCTIS-1}
\end{equation}
The dispersion contour for this PCTIS is shown in Fig. \ref{Fig:PCTIS beamshifter_dispersoin}.

\begin{figure}
\begin{centering}
\includegraphics[width=3in]{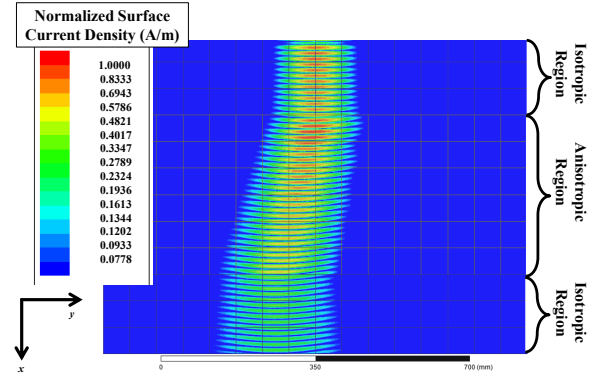}
\par\end{centering}
\caption{Normalized surface current density for the PCTIS beam-shifting surface.
The incoming beam is deflected by $-13.93^{\circ}$ in the anisotropic
region. The total size of the surface is $96\times72$ cm ($32\lambda_0\times24\lambda_0$). Each isotropic region
is $96\times18$ cm ($32\lambda_0\times6\lambda_0$). The dimensions of the anisotropic region are $96\times36$ cm ($32\lambda_0\times12\lambda_0$). }
\label{Fig:PCTIS beamshifter}
\end{figure}

The beam-shifter was simulated using HFSS. The isotropic and
anisotropic regions were modeled using the screening impedance boundary condition over a grounded dielectric substrate.
The boundaries of the simulation domain were terminated with radiation
boundaries, and one edge was illuminated with a Gaussian beam. The
results of the simulation at 10 GHz are shown in Fig. \ref{Fig:PCTIS beamshifter}.
As expected, the Gaussian beam is refracted by
$-13.93^{\circ}$ upon encountering the anisotropic medium.

\begin{figure}
\begin{centering}
\includegraphics[width=3in]{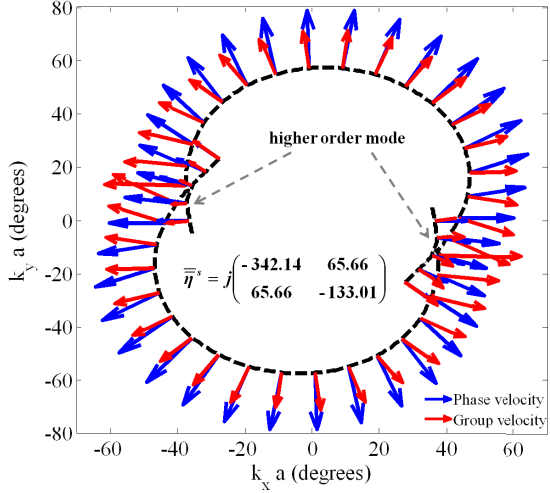}
\par\end{centering}
\caption{The isofrequency dispersion contour at 10 GHz for the anisotropic surface of a PCTIS
beam-shifter with two modes excited.}
\label{Fig:PCTIS multiple modes}
\end{figure}
\begin{figure}
\begin{centering}
\includegraphics[width=3in]{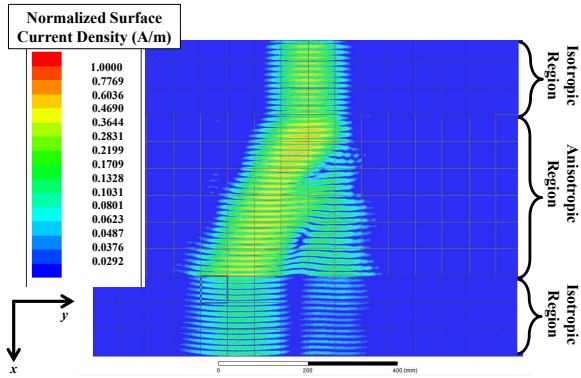}
\par\end{centering}
\caption{Normalized surface current density for the PCTIS beam-shifting surface.
The incoming beam is deflected in two different directions in the
anisotropic region. This is due to the presence of two modes as shown
in Fig. \ref{Fig:PCTIS beamshifter_dispersoin}. The power in the $TM$ mode is refracted by $-22.42^\circ$ and the power in the higher order $TE$ mode is refracted by $7.78^\circ$. The total size of the surface is $96\times72$ cm ($32\lambda_0\times24\lambda_0$). Each isotropic region
is $96\times18$ cm ($32\lambda_0\times6\lambda_0$). The dimensions of the anisotropic region are $96\times36$ cm ($32\lambda_0\times12\lambda_0$).}
\label{Fig:PCTIS multiple modes beamsplit}
\end{figure}

For the chosen substrate, $Y_{critical}^{s}=1/(-129.72j)$ S at 10 GHz.
The eigenvalues (diagonalized sheet admittance values) of \eqref{eq:anisotropic region PCTIS-1}
are $Y_{\lambda_{1}}^{s}=1/(-301.22j)$ S and $Y_{\lambda_{2}}^{s}=1/(-136.25j)$
S, and therefore satisfy \eqref{eq:cond1} and \eqref{eq:cond2}.
If conditions \eqref{eq:cond1} and \eqref{eq:cond2} were not satisfied,
a $TM$ and a $TE$ mode would co-exist. The 10 GHz dispersion contour for such
a situation is shown in Fig. \ref{Fig:PCTIS multiple modes}. The
sheet impedance corresponding to this dispersion contour is
\begin{equation}
\eta_{sheet=}^{\prime\prime}(Y_{sheet}^{\prime\prime})^{-1}=j\begin{pmatrix}-342.14 & 65.66\\
65.66 & -133.01
\end{pmatrix}\Omega.\label{eq:anisotropic region PCTIS-2}
\end{equation}
and the eigenvalues of $Y_{sheet}^{\prime\prime}$ are $Y_{\lambda_{1}}^{s}=1/(-114.11j)$
S and $Y_{\lambda_{2}}^{s}=1/(-361.04j)$ S. In this case, propagation
along certain directions of the the beam-shifter will produce two beams.
This is verified with full-wave simulation (results shown in Fig. \ref{Fig:PCTIS multiple modes beamsplit})
for propagation along the $x$-axis.

\subsection{Realization}

\begin{figure}
\begin{centering}
\includegraphics[width=2in]{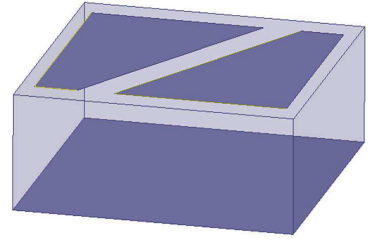}
\par\end{centering}
\caption{Possible choice for unit cell \cite{Sievenpiper_Journal} for PCTIS beam-shifter implementation (anisotropic region).
Dark areas represent metal. The sheet impedance can be designed to be identical to \eqref{eq:anisotropic region PCTIS-2}. Other possible unit cell geometries  based on circular or elliptical patches can also be used \cite{Maci_Isoflux}.}
\label{Fig:PCTIS unit cell geometry}.
\end{figure}

The PCTIS beam-shifter of Fig. \ref{Fig:PCTIS beamshifter} can be implemented
by patterning the metallic cladding above a $1.27$ mm thick grounded
dielectric substrate with $\epsilon_{r}=10.2$. Using the sheet extraction
method described in \cite{Patel_Grbic_Analytical_journal}, a unit cell can be designed for the anisotropic
region (see Fig. \ref{Fig:PCTIS unit cell geometry}) that has a sheet impedance identical to that of \eqref{eq:anisotropic region PCTIS-2}.
The isotropic region of the beam-shifter can be implemented by printing
a square patch over the grounded dielectric substrate, similar to Fig. \ref{Fig:PCTIS unit cell geometry}, but with the diagonal gap removed.

\section{Conclusion}

In this paper, a method for designing transformation electromagnetics devices using tensor impedance surfaces (TISs) was presented. It was shown that transforming an idealized tensor impedance boundary condition (TIBC) according to the transformation electromagnetics method, results in a transformation of the free space above it.
An alternate method was proposed that allows transformation electromagnetics devices to be implemented using TIBCs, while maintaining free space above. The procedure was extended to include printed-circuit tensor impedance surfaces (PCTISs), which are practical realizations of TIBCs, and consist of a patterned metallic cladding over a grounded dielectric substrate. The alternate method allows anisotropic TIBCs and PCTISs to be designed that support tangential wave vector and Poynting vector distributions specified by a coordinate transformation. Beam-shifters are designed (both a TIBC and a PCTIS version) that laterally shift a surface wave beam at 10 GHz. The design methods reported in this paper may be applicable to graphene-based devices since an infinitesimally thin graphene sheet can be characterized by a conductivity tensor, using a non-local model for graphene \cite{Hanson_aniostropic}. Preliminary results of this work were presented at the 2013 IEEE International Microwave Symposium \cite{Patel_IMS2013_transformation}.

\ifCLASSOPTIONcaptionsoff \newpage{}\fi

\vfill{}








\end{document}